\documentclass[12pt,preprint]{aastex}

\begin{document}

\title{CORRELATIONS AMONG JET, ACCRETION DISK, AND BROAD LINE REGION OF FLAT SPECTRUM RADIO QUASARS}
\author{Jin Zhang\altaffilmark{1,2}, Zi-Wei Xue\altaffilmark{1,5}, Jian-Jian He\altaffilmark{4,5}, En-Wei Liang\altaffilmark{1,2,3}, Shuang-Nan Zhang\altaffilmark{1,4,6}}
\altaffiltext{1}{Key Laboratory of Space Astronomy and Technology, National Astronomical Observatories, Chinese Academy of Sciences, Beijing 100012, China; zhang.jin@hotmail.com; zhangsn@ihep.ac.cn}
\altaffiltext{2}{Guangxi Key Laboratory for Relativistic Astrophysics, Guangxi University, Nanning 530004, China}
\altaffiltext{3}{Department of Physics, Guangxi University, Nanning 530004, China}
\altaffiltext{4}{Key Laboratory of Particle Astrophysics, Institute of High Energy Physics, Chinese Academy of Sciences, Beijing 100049, China}
\altaffiltext{5}{University of Chinese Academy of Sciences, Beijing 100049, China}
\altaffiltext{6}{Physics Department, University of Alabama in Huntsville, Huntsville, AL 35899, USA}
\begin{abstract}

The broadband spectral energy distributions (SEDs) of 18 GeV flat spectrum radio quasars (FSRQs) are collected and compiled from literature, in which both the jet emission and the accretion disk radiation can be observed, in order to investigate the correlations among their jet power ($P_{\rm jet}$), accretion disk luminosity ($L_{\rm disk}$), and luminosity of broad line region (BLR, $L_{\rm BLR}$). On the basis of the SED fits with the jet radiation and accretion disk radiation models, we calculate $P_{\rm jet}$ and $L_{\rm disk}$. No correlation between $P_{\rm jet}$ with either $L_{\rm disk}$ or $L_{\rm BLR}$ is found. With a sub-sample of $L_{\rm BLR}$ for 13 GeV FSRQs, it is observed that $L_{\rm BLR}$ is strongly correlated with their $L_{\rm disk}$. We also study the BLR covering factors ($L_{\rm BLR}/L_{\rm disk}$) of the GeV FSRQs in our sample, averagely which are smaller than that of the large samples of radio-loud and radio-quiet quasars. $P_{\rm jet}$ of some GeV FSRQs is higher than $L_{\rm disk}$, but $P_{\rm jet}$ of all the GeV FSRQs is lower than the accretion power of black hole (BH), which is estimated by $\dot{M}c^2=L_{\rm disk}/0.1$, indicating that the total accretion power of BH is sufficient to drive the jets in these sources; however the uncorrelation between $L_{\rm disk}$ and $P_{\rm jet}$ of the GeV FSRQs may suggest that their jets are launched by the Blandford$-$Znajek process via extracting the rotational energy of BH. Using the $L_{\rm BLR}$--$L_{\rm disk}$ relation of the GeV FSRQs, we estimate $L_{\rm disk}$ of a BL Lac sample with their $L_{\rm BLR}$. A comparison of $L_{\rm BLR}$ and Eddington ratio ($L_{\rm disk}/L_{\rm Edd}$) among BL Lacs, very radio-loud NLS1 galaxies, and FSRQs is also presented. It is found that along with the BL Lac--NLS1--FSRQ sequence $L_{\rm BLR}$ and $L_{\rm disk}/L_{\rm Edd}$ increase, which may correspond to the change of the accretion disk structure and the transformation of the dominant mechanism for jet launching. This is also consistent with the division of their parent populations, i.e., low-excitation radio galaxies and high-excitation radio galaxies.

\end{abstract}


\section{Introduction}           
\label{sect:intro}

It is generally believed that there is a supermassive black hole (BH) at the center of an active galactic nucleus (AGN); the gravitational potential of the BH is the ultimate energy source of the AGN. The accreted matter of a BH forms an accretion disk and loses angular momentum through viscous or turbulent processes (Rees 1984). Broad emission lines are one of the dominant features of many AGN spectra, which are produced in a `broad line region' (BLR) nearby the BH. The BLR is widely acknowledged to be photoionized by the continuum from the accretion disk (e.g., Kwan \& Krolik 1981; Ferland et al. 1992). In order to explain the special radiation properties of BL Lac objects (BL Lacs), Blandford \& Rees (1978) proposed that there are relativistic jet structures in these sources. The jet launching is reported to be connected with the central BH, accretion disk, and corona in a source (Armitage \& Natarajan 1999; Merloni \& Fabian 2002; Wang et al. 2004; Cao 2004, 2014; Yuan \& Narayan 2014; Wilkins \& Gallo 2015; Chen et al. 2015). The jets may be powered via the Blandford$-$Payne (BP; Blandford \& Payne 1982) process by releasing the gravitational energy of accreting matter that moves toward the BH, and/or via Blandford$-$Znajek (BZ; Blandford \& Znajek 1977) mechanism by extracting the rotational energy of BH.

So far, most confirmed extragalactic GeV-TeV emitters are blazars, a sub-sample of radio-loud AGNs with relativistic jets. Blazars are divided into BL Lacs and flat spectrum radio quasars (FSRQs) according to their emission line features. Generally, the broadband spectral energy distributions (SEDs) of BL Lacs can be well explained with the synchrotron radiation and the synchrotron self-Compton (SSC) scattering of the relativistic electrons in jets (Maraschi et al. 1992; Ghisellini et al. 1996, 2010; Zhang et al. 2012; Liao et al. 2014; Yan et al. 2014); however, FSRQs need a black body spectrum component and a component from inverse Compton (IC) scattering of the external photons to fit their broadband SEDs (e.g., Sikora et al. 1994; Ghisellini et al. 1996, 2010; Zhang et al. 2014; Cao \& Wang 2013; Chen et al. 2010, 2012). Compared with BL Lacs, FSRQs have prominent emission lines in their spectra, and sometimes there are significant blue bumps at $\sim10^{15}$ Hz in their observed broadband SEDs, which is believed to be the thermal emission of accretion disk. However, often no significant blue bump is observed in the SEDs since the nonthermal radiation from jet overwhelms the thermal mission from accretion disk, especially when the sources are in the high state (e.g., for 3C 454.3, Jorstad et al. 2013). Hence, the GeV FSRQs in which the broadband SEDs are dominated by the jet emission and the significant blue bumps of the thermal emission from accretion disk are observed in their SEDs, are the good candidates for investigating the correlations among their jet power, accretion disk radiation, and BLR luminosity.

The flux ratios of emission lines from BLR are suggested to be approximately constant among different AGNs (Francis et al. 1991; Vanden Berk et al. 2001). If setting the emission line luminosity of Ly$\alpha$ to 100, the relative luminosities of H$\beta$, H$\alpha$, Mg~{\scriptsize II}, and C~{\scriptsize IV} lines are 22, 77, 34, and 63, respectively, and then the total emission line luminosity is 555.8 (Francis et al. 1991; Celotti et al. 1997). Therefore, the luminosity of BLR can be roughly estimated with the luminosities of one or some emission lines. Using the continuum luminosity and the emission line luminosity in Shen et al. (2011), Calderone et al. (2013) calculated $L_{\rm disk}$ and $L_{\rm BLR}$, respectively, and reported that $\log (L_{\rm disk}/L_{\rm BLR})\sim0.92\pm0.28-1.08\pm0.28$, i.e., the covering factor of BLR is probably equal to 10 percent; therefore, the value of $10L_{\rm BLR}$ is always used as a proxy to the accretion disk luminosity ($L_{\rm disk}$). Sometimes $0.1L_{\rm disk}$ is also used to estimate the luminosity of BLR in blazars when authors have obtained $L_{\rm disk}$ and want to calculate the energy density of BLR (e.g., Ghisellini et al. 2010; Zhang et al. 2014). However, so far it has not be examined whether the $L_{\rm BLR}/L_{\rm disk}\sim0.1$ relation is still valid for blazars, especially for the GeV--TeV blazars.

In this paper, a sample of SEDs for GeV FSRQs is compiled to study their jet and accretion disk relations and their covering factors of BLRs. The sample selection and data are presented in Section 2. The models and results of SED fitting are described in Section 3. The study of the covering factors of BLRs is in Section 4. The possible connection between the jet and accretion disk is shown in Section 5. Further investigation to the correlations among the jet, BLR, and accretion disk is described in Section 6. A simple discussion on the anisotropic radiation of accretion disk, the jet formation mechanism for GeV AGNs, and the implications of unification for radio loud AGNs is given in Section 7. A summary for our results is in Section 8. Throughout, $H_{0}$=73 km s$^{-1}$ Mpc$^{-1}$, $\Omega_{\Lambda}=0.73$ and
$\Omega_{m}=0.27$ are adopted.

\section{Sample and Data}

\subsection{Sample Selection and SED Data}

\begin{figure*}
\includegraphics[angle=0,scale=0.1]{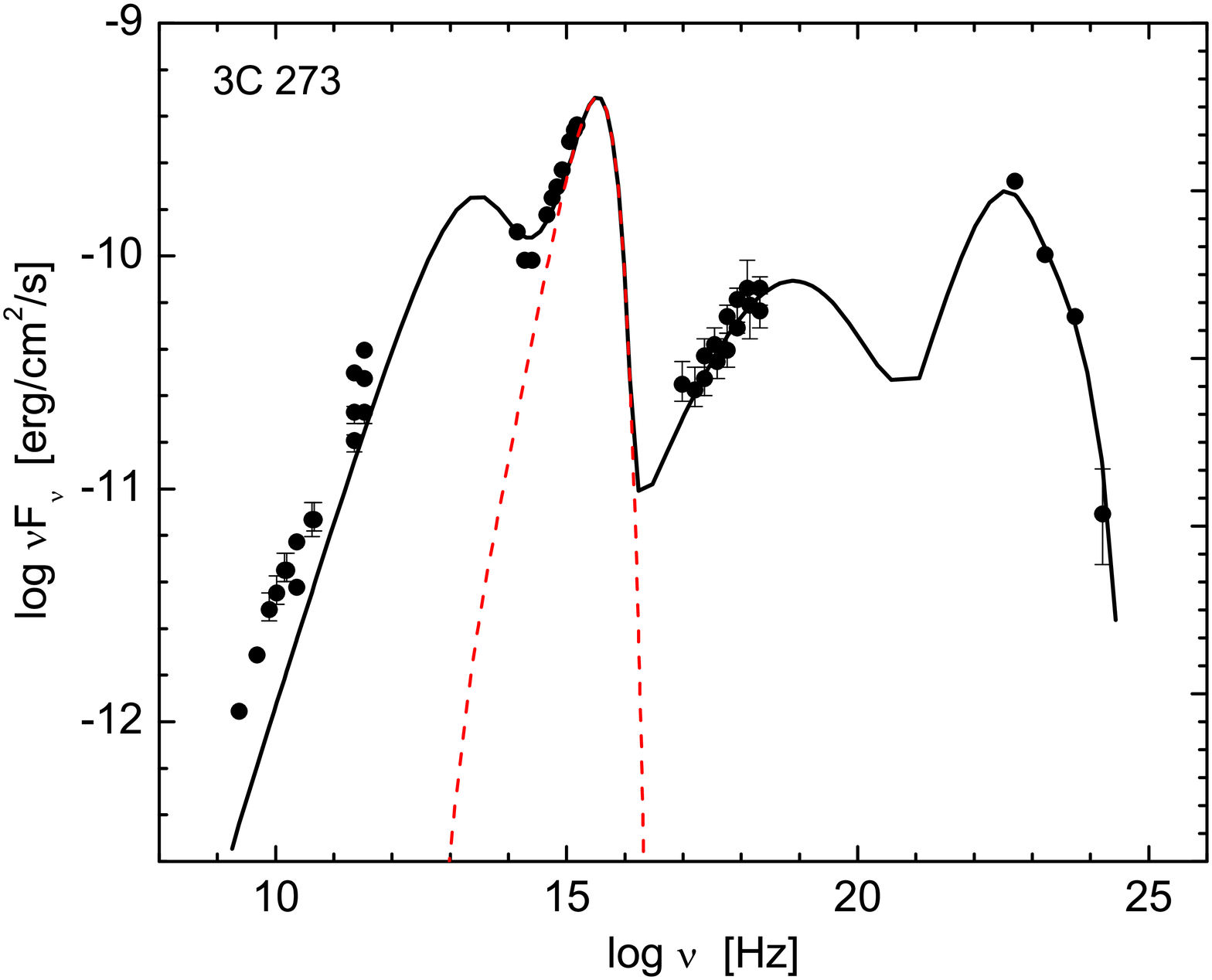}
\includegraphics[angle=0,scale=0.1]{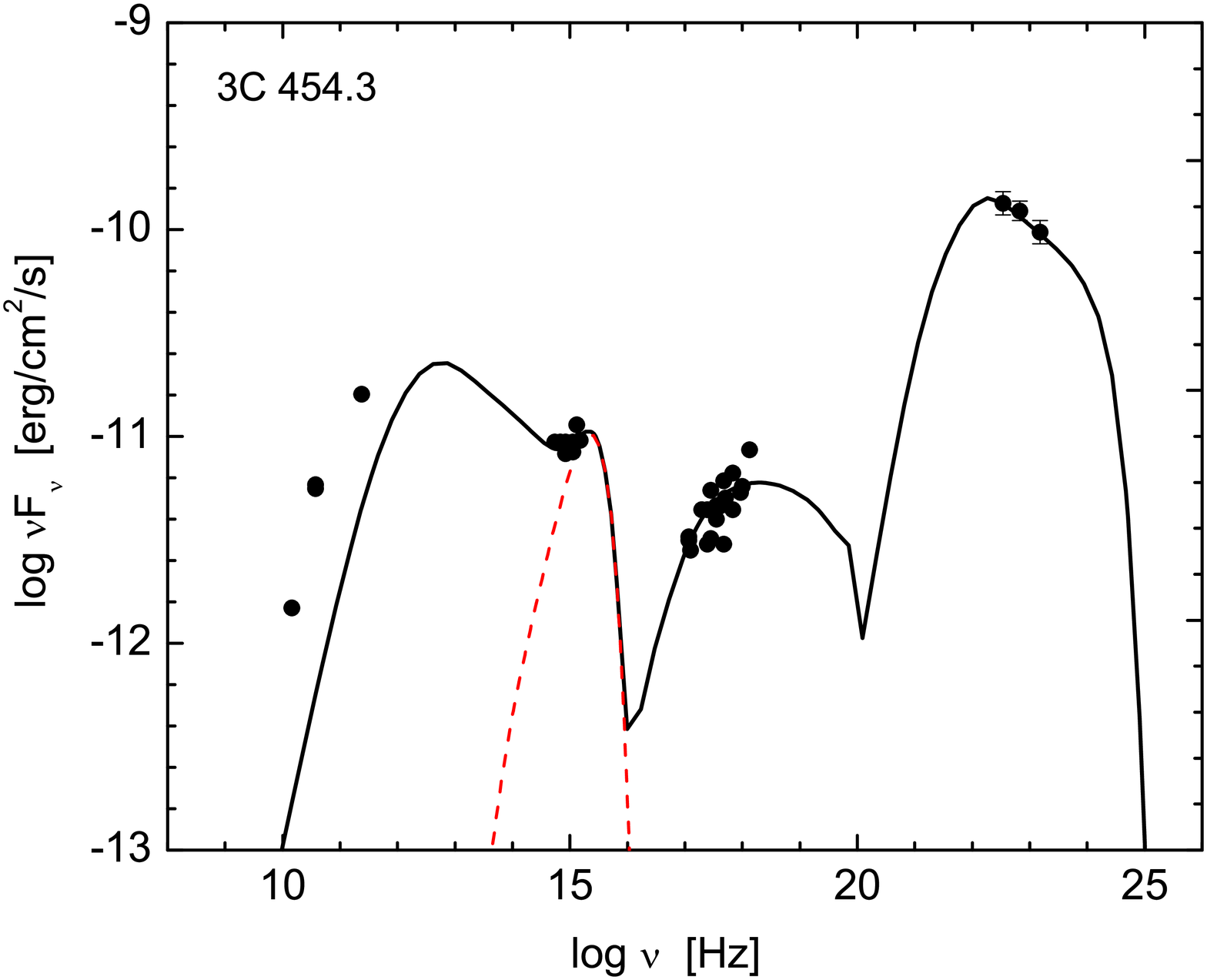}
\includegraphics[angle=0,scale=0.1]{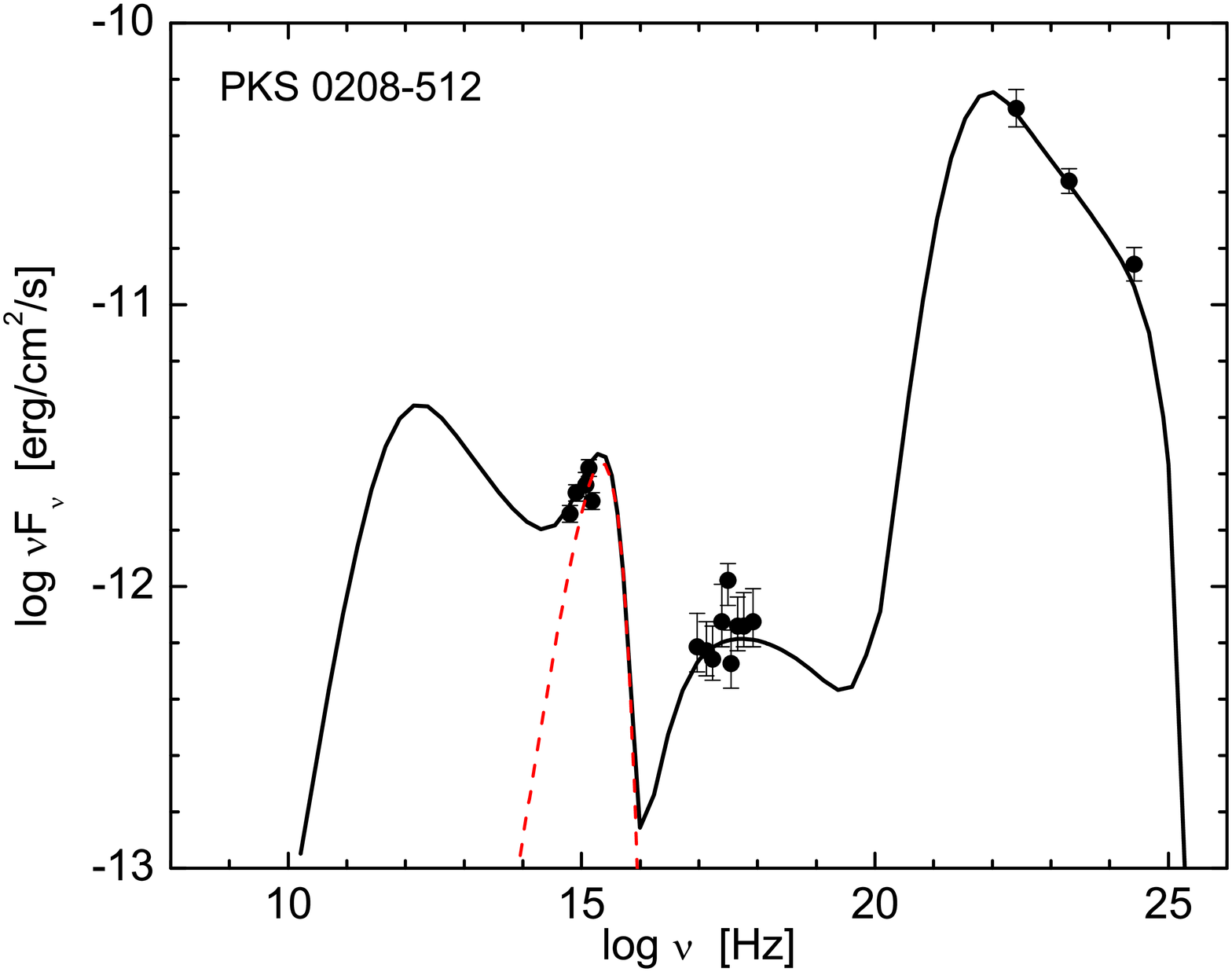}
\includegraphics[angle=0,scale=0.1]{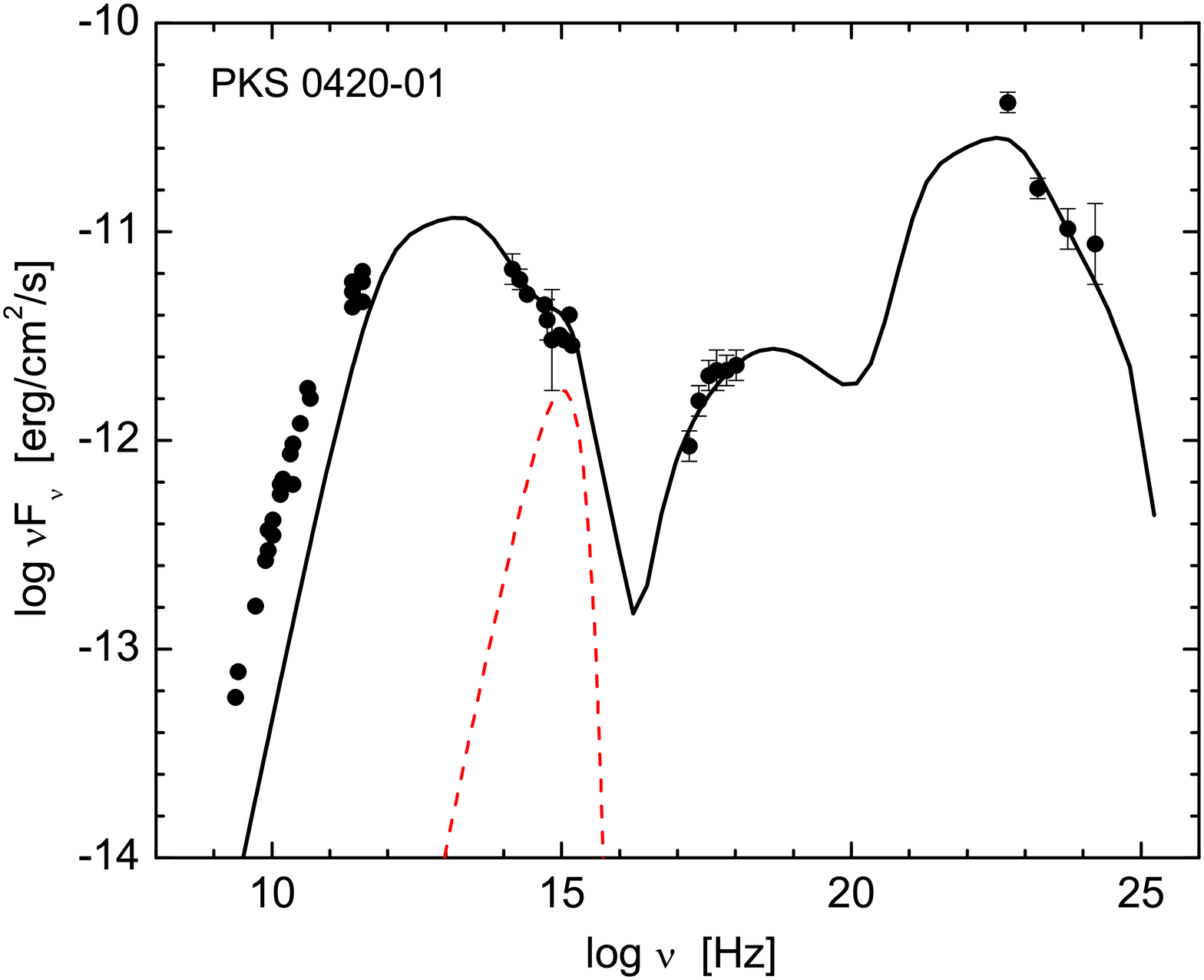}
\includegraphics[angle=0,scale=0.1]{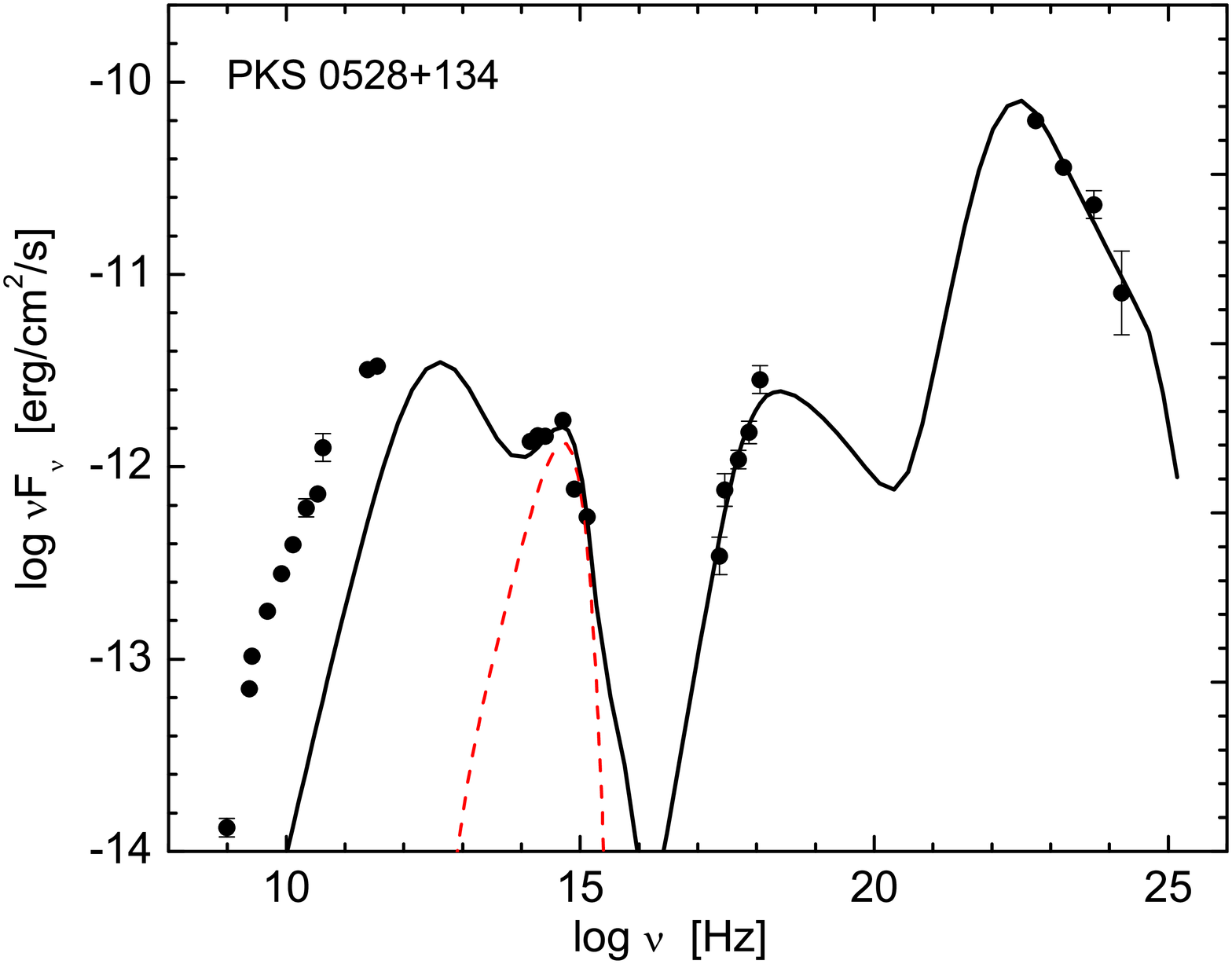}\\
\includegraphics[angle=0,scale=0.1]{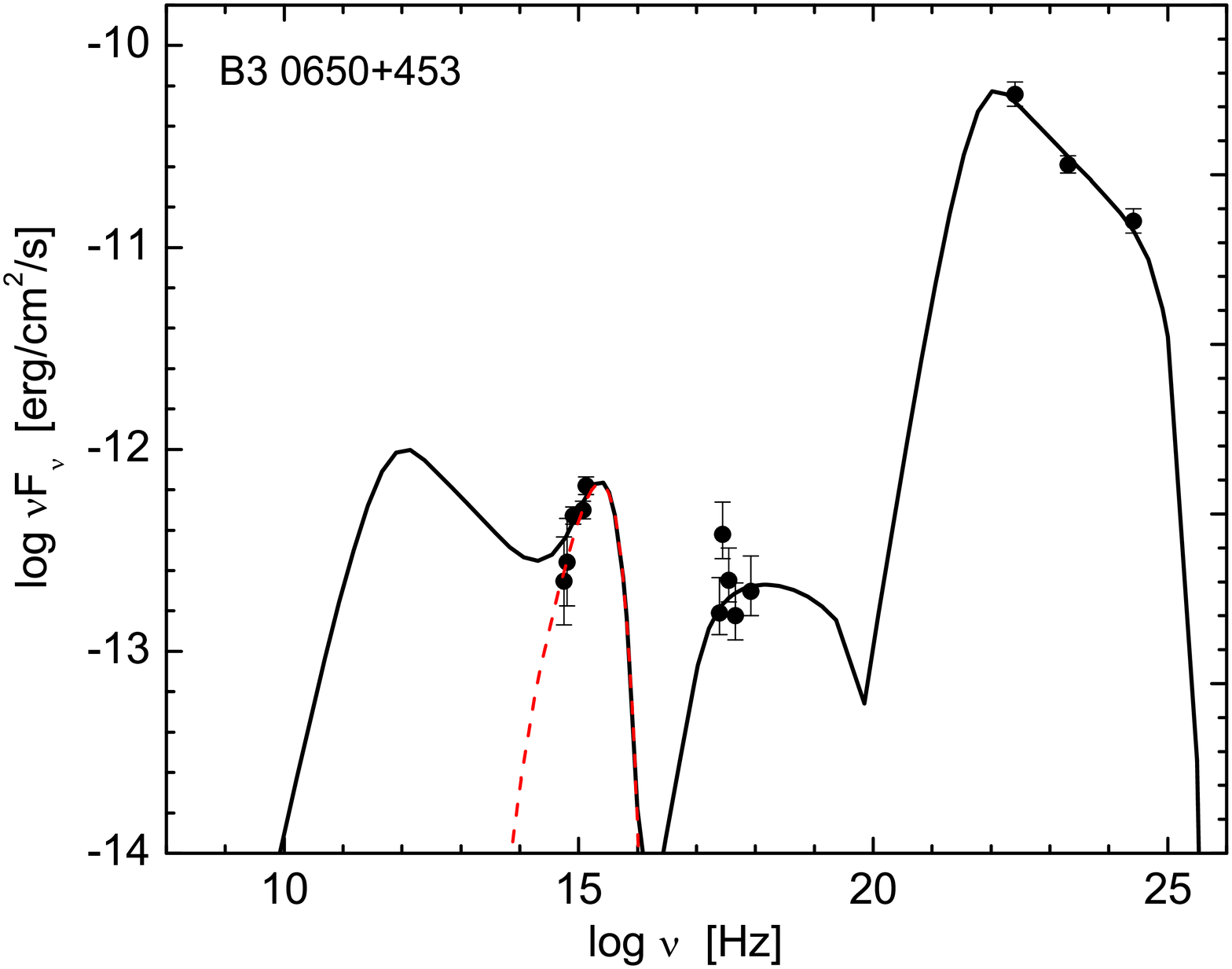}
\includegraphics[angle=0,scale=0.1]{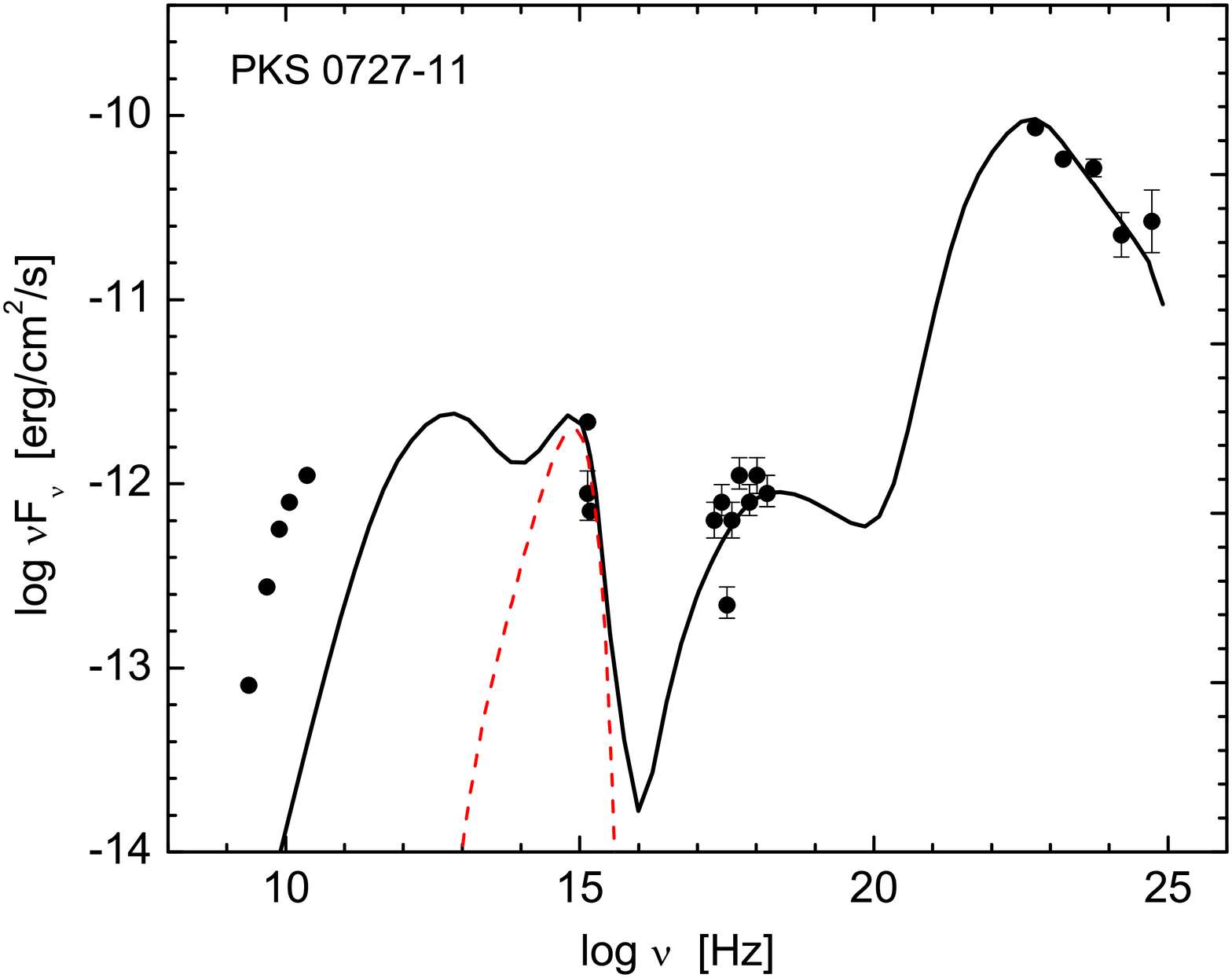}
\includegraphics[angle=0,scale=0.1]{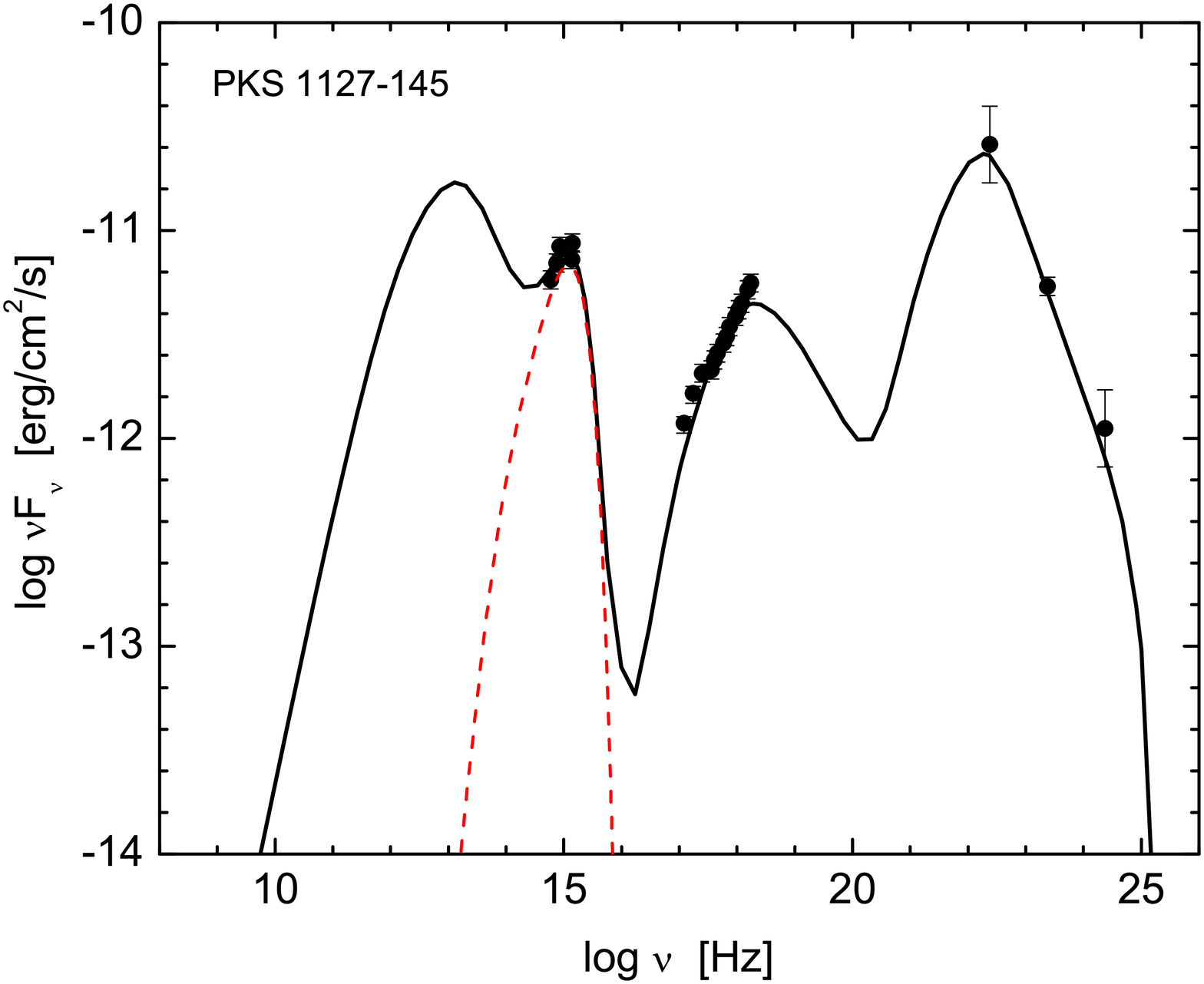}
\includegraphics[angle=0,scale=0.1]{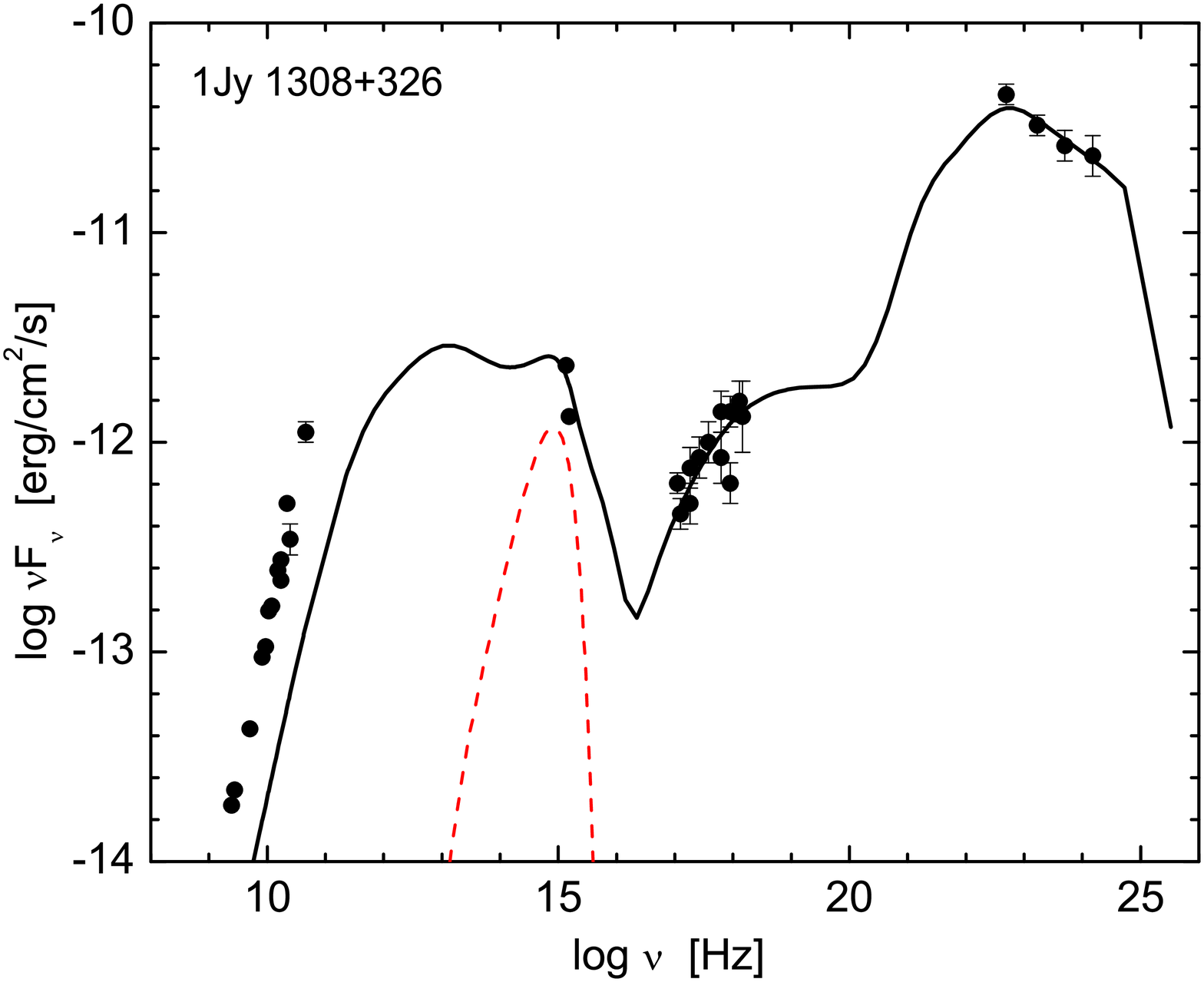}
\includegraphics[angle=0,scale=0.1]{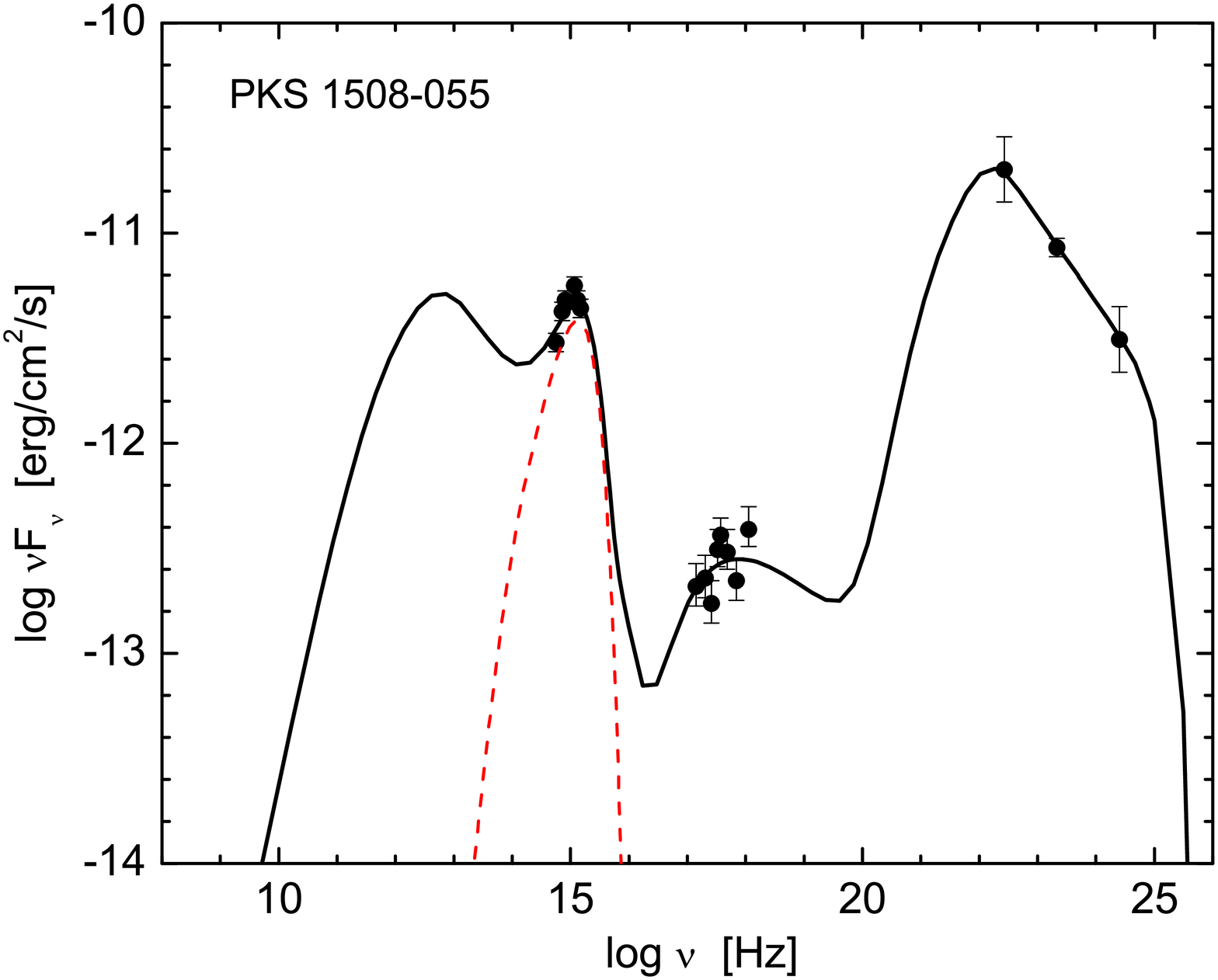}\\
\includegraphics[angle=0,scale=0.1]{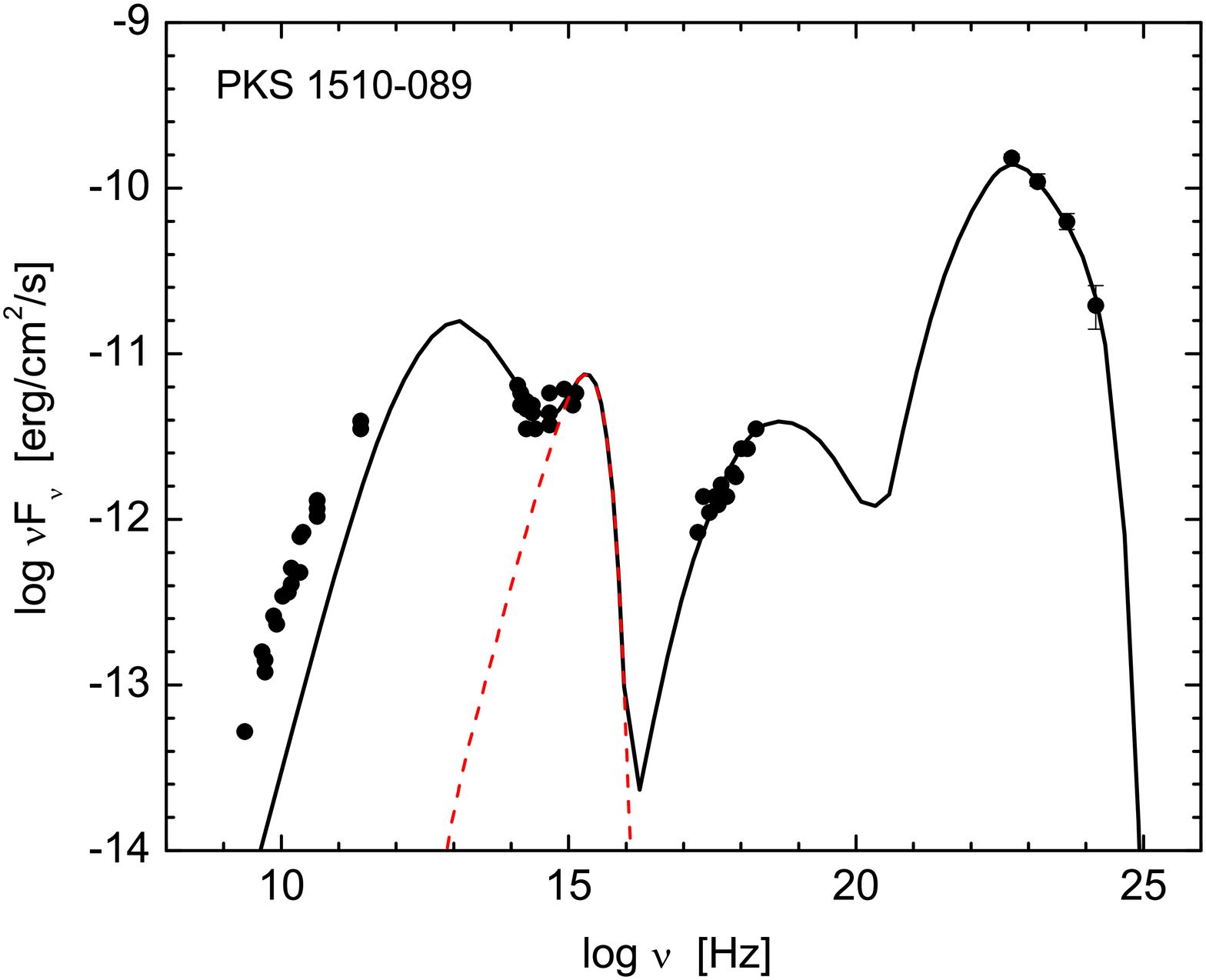}
\includegraphics[angle=0,scale=0.1]{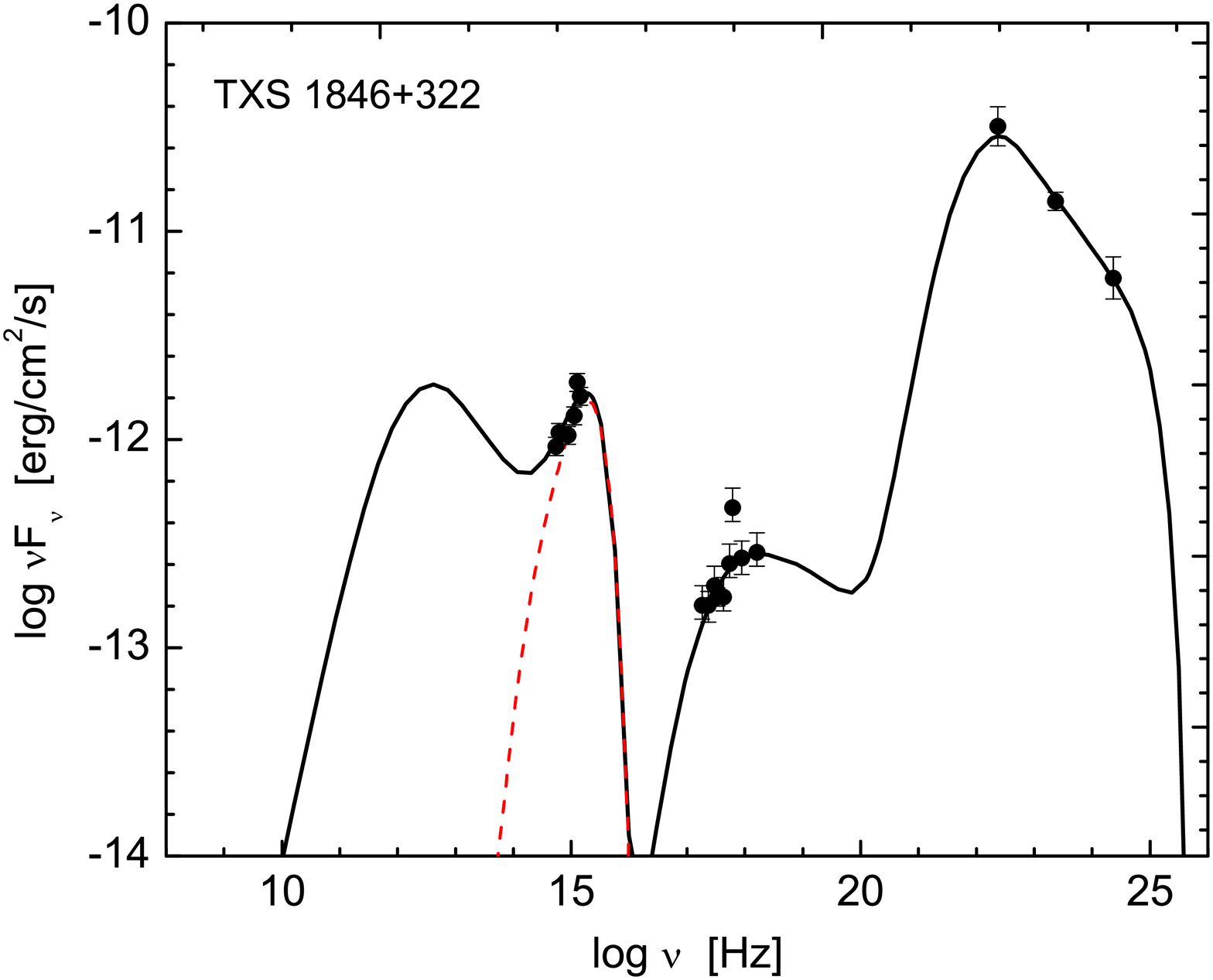}
\includegraphics[angle=0,scale=0.1]{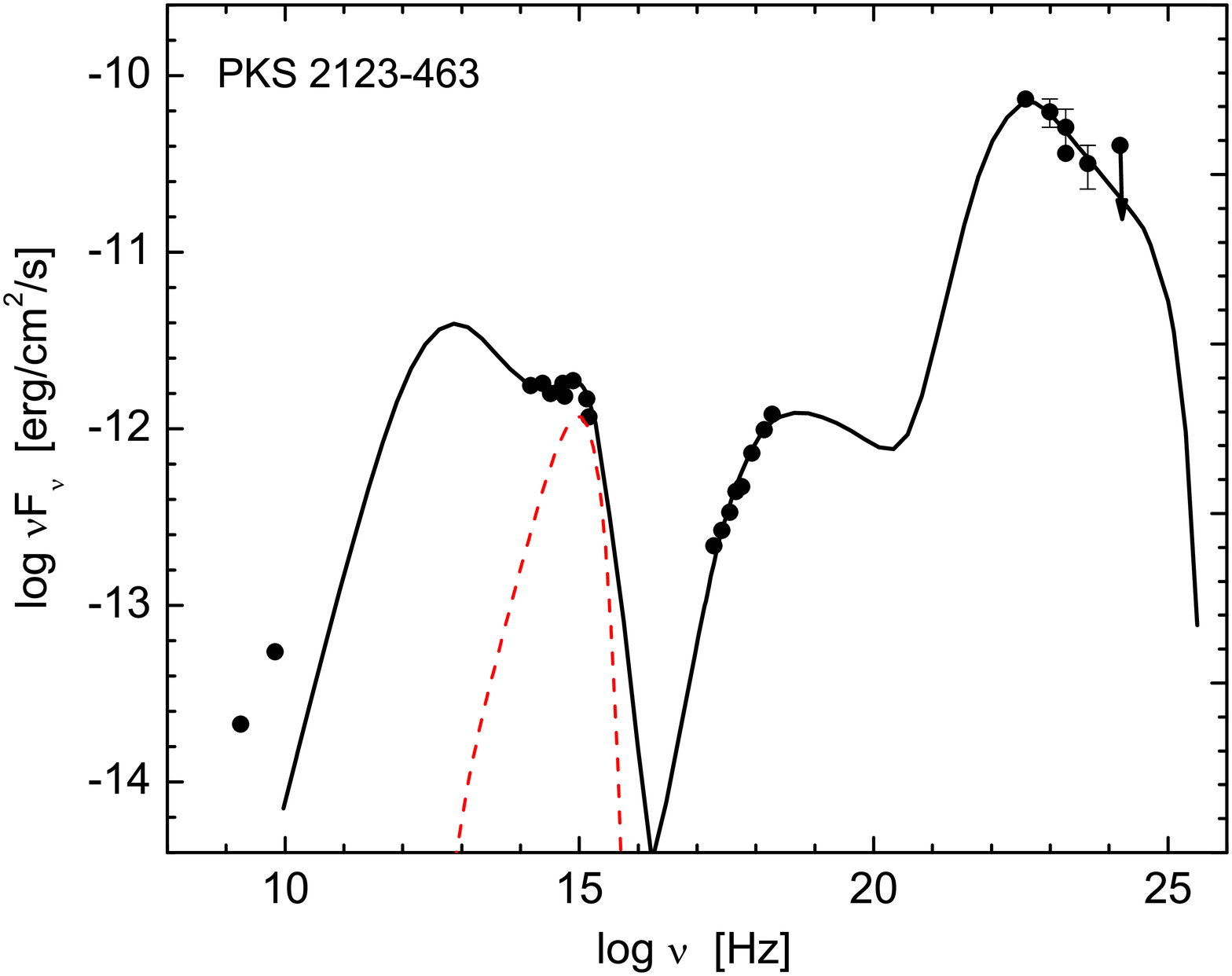}
\includegraphics[angle=0,scale=0.1]{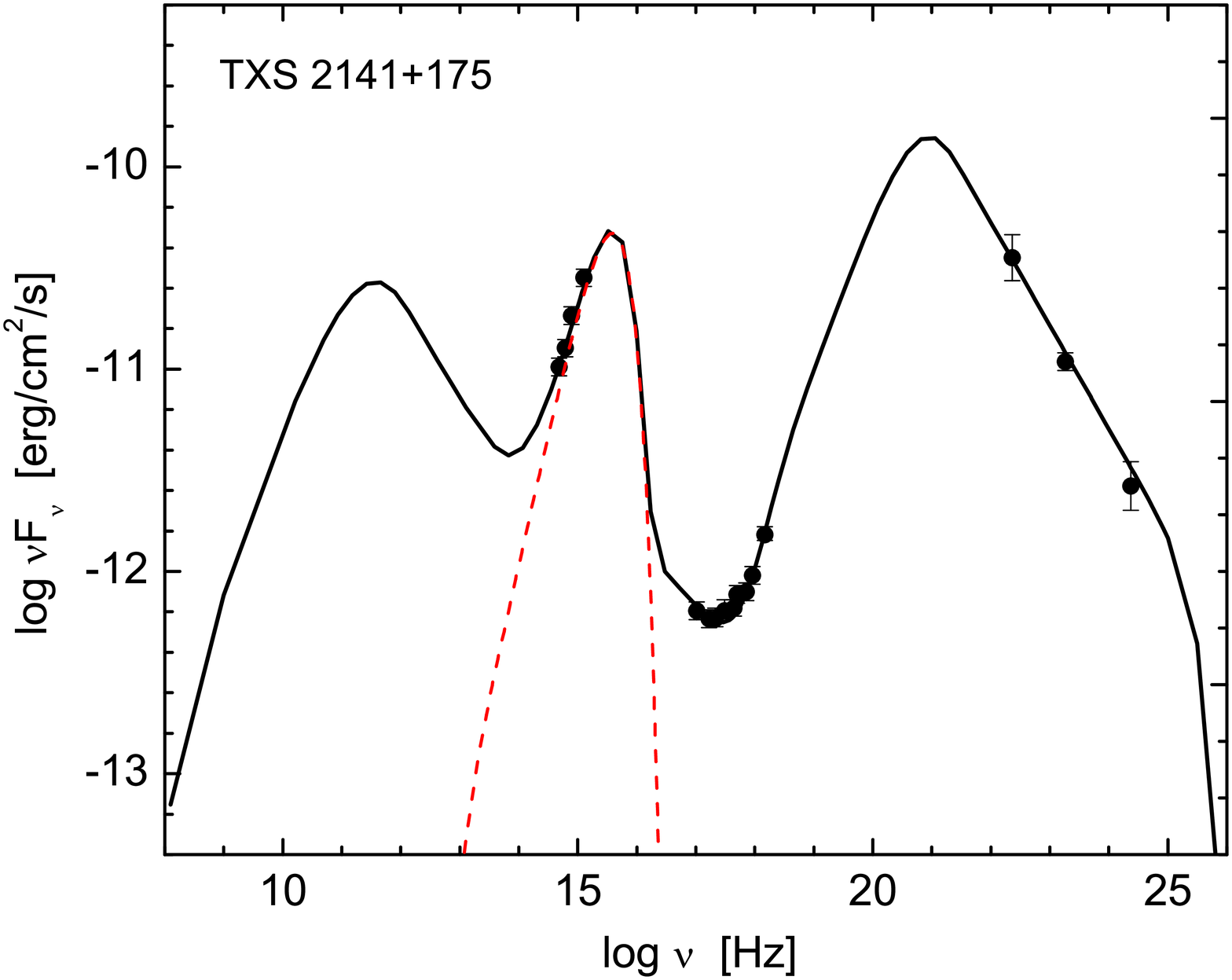}
\includegraphics[angle=0,scale=0.1]{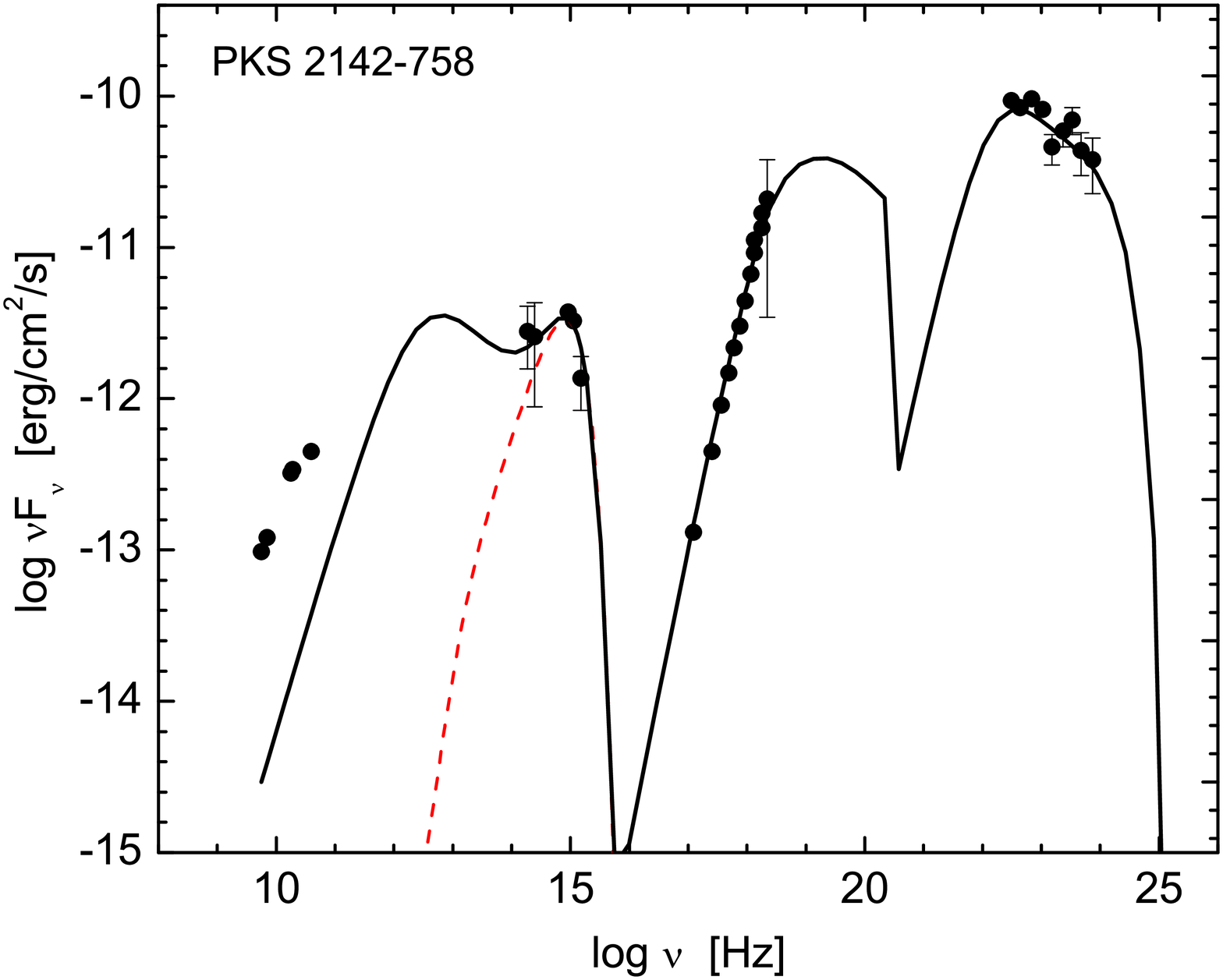}\\
\includegraphics[angle=0,scale=0.1]{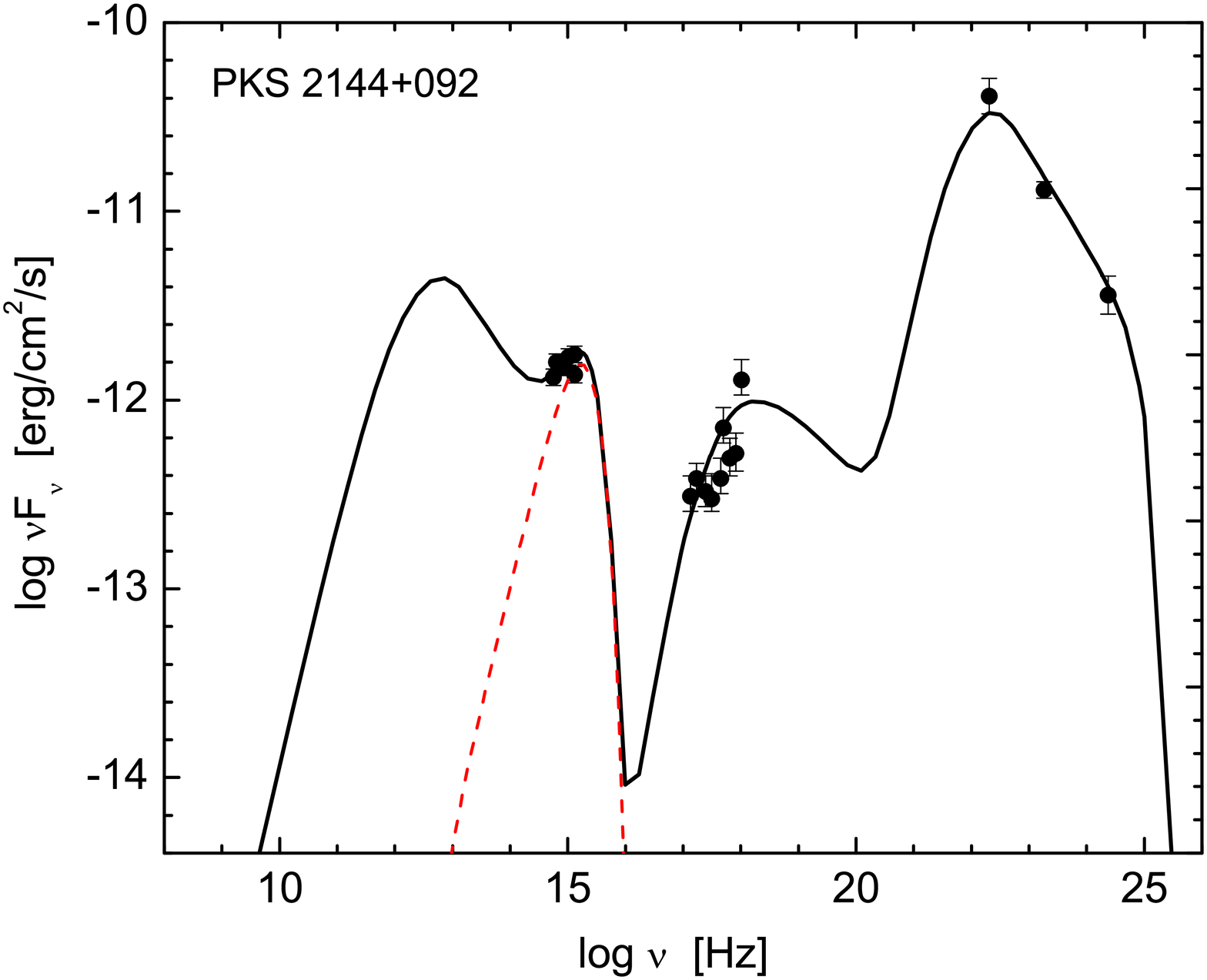}
\includegraphics[angle=0,scale=0.1]{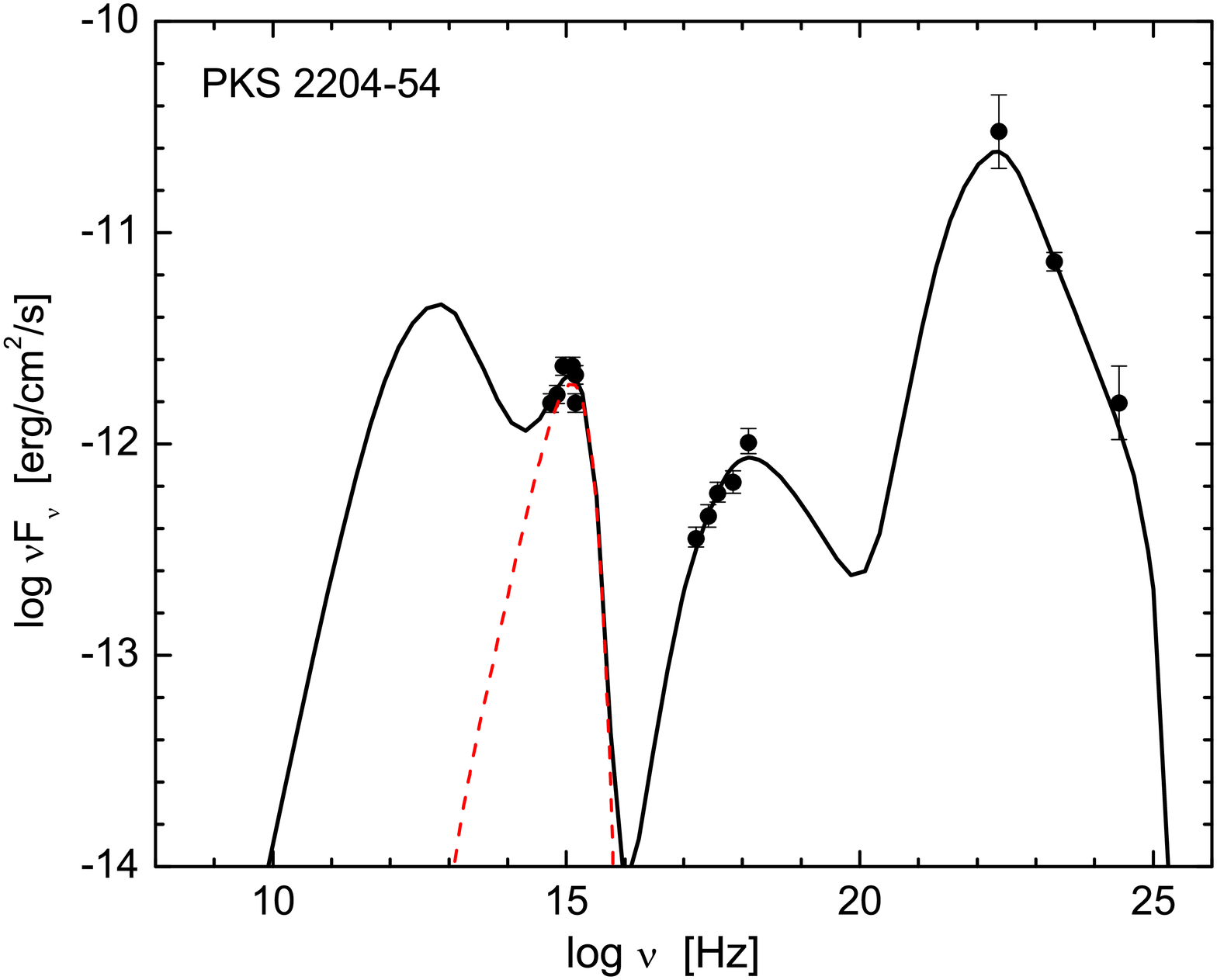}
\includegraphics[angle=0,scale=0.1]{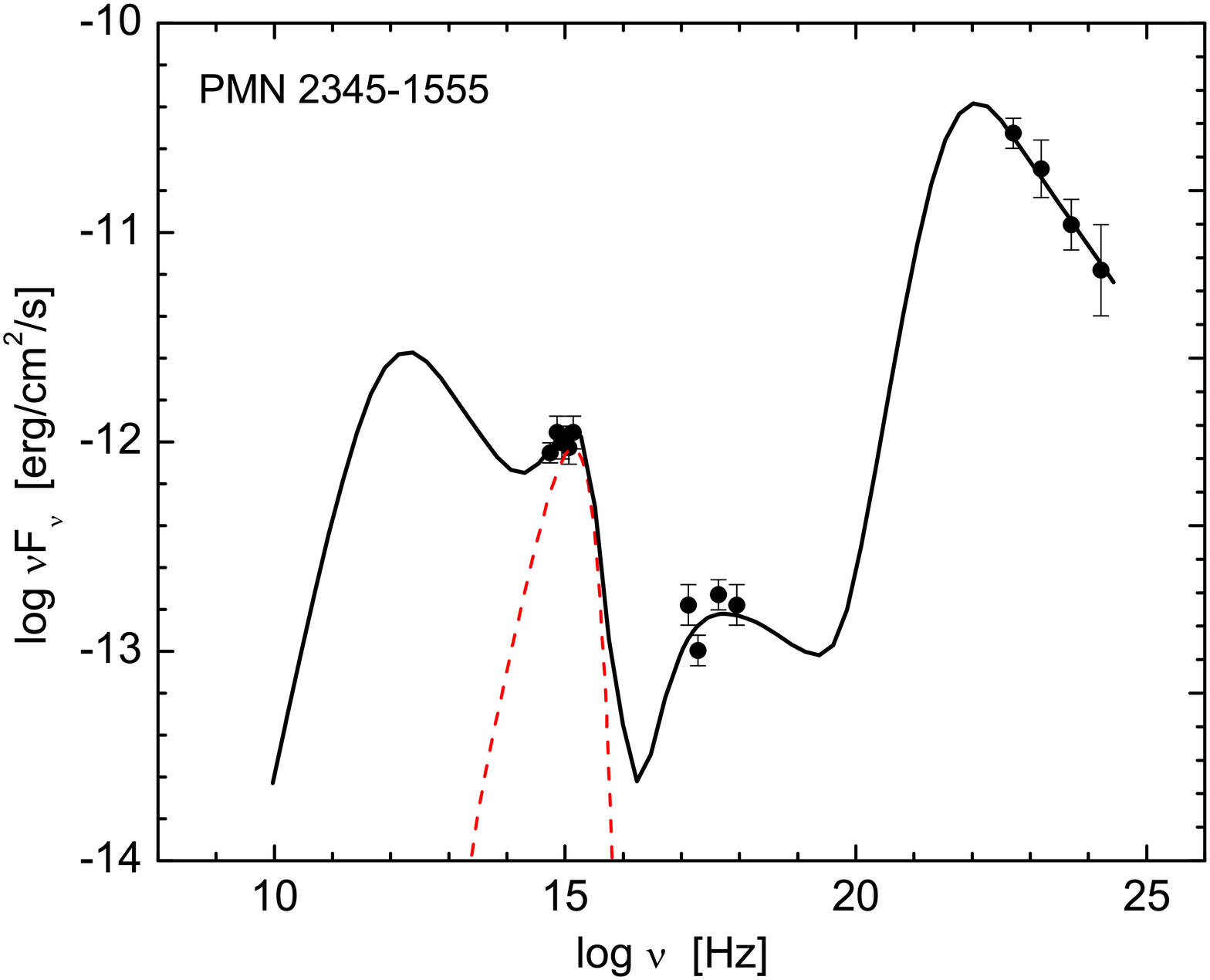}
\caption{Observed SEDs (black data points) with the model best fits (lines) for the 18 GeV-FSRQs in our sample. The black solid lines are the sum of emission from the jet and the accretion disk (red dashed lines).} \label{SED}
\end{figure*}

We select the \emph{Fermi}/LAT FSRQs with confirmed redshift, simultaneously or quasi-simultaneously observed broadband SEDs, and obvious blue bumps in their SEDs. There are 18 sources (hereafter GeV-FSRQs) as shown in Figure \ref{SED}; seven GeV-FSRQs (3C 273, PKS 0420$-$01, PKS 0528+134, PKS 0727$-$11, 1Jy 1308+326, PKS 1510$-$089, PMN 2345$-$1555) from Zhang et al. (2014), eight GeV-FSRQs (PKS 0208$-$512, B3 0650+453, PKS 1127$-$145, PKS 1508$-$055, TXS 1846+322, TXS 2141+175, PKS 2144+092, PKS 2204$-$54) from Ghisellini et al. (2010), and three GeV-FSRQs from other literature (3C 454.3 in Vercellone et al. 2012; PKS 2123$-$463 in D'Ammando et al. 2012; PKS 2142$-$758 in Dutka et al. 2012).

\subsection{Luminosity and Energy Density of BLR}

The BLR luminosity ($L_{\rm BLR}$) is calculated with the method proposed by Celotti et al. (1997) using the relative weights of different emission lines and the fluxes of emission lines whose observation data are available. The BLR luminosities of GeV-FSRQs in our sample are taken from Xiong \& Zhang (2014), except for PKS 0528+134, for which $L_{\rm BLR}$ is calculated using the flux of emission line C~{\scriptsize IV} and Equation (1) in Celotti et al. (1997). If there are more than one value available for $L_{\rm BLR}$ in Xiong \& Zhang (2014), the average of $L_{\rm BLR}$ is used in this paper. A sub-sample of $L_{\rm BLR}$ for 13 GeV-FSRQs in our sample is obtained, and the values of $L_{\rm BLR}$ are shown in Table 1. We also consider the correction of $L_{\rm BLR}$ since the different cosmological parameters are used in those works.

The energy density of BLR at rest frame is calculated by $U_{\rm BLR}=\frac{L_{\rm BLR}}{4\pi R_{\rm BLR}^2 c}$, where $R_{\rm BLR}$ is the radius of BLR. The radius of BLR is estimated by the relations of $\log R_{\rm BLR}=-21.3^{+2.9}_{-2.8}+0.519^{+0.063}_{-0.066}\log L_{5100}/{10^{44} \rm erg~ s^{-1}}$ ($R_{\rm BLR}$ in units of light days, Bentz et al. 2009) and $L_{\rm H\beta}=(1.425\pm0.007)\times10^{42}(\frac{L_{5100}}{10^{44}\rm erg~s^{-1}})^{1.133\pm0.005} ~\rm erg~s^{-1}$ (Greene \& Ho 2005), where $L_{5100}$ and $L_{\rm H\beta}$ are the luminosities of continuum at $5100~{\rm\AA}$ and $\rm H\beta$ line, respectively. If there is no observational flux of $\rm H\beta$ line available, $L_{\rm H\beta}$ is derived by the luminosities of other emission lines or $L_{\rm BLR}$ using the relative flux ratio among different emission lines (Francis et al. 1991; Celotti et al. 1997); please refer to Table 1 for more details.

\subsection{Black Hole Mass}

The mass of BH, along with mass accretion rate, is a fundamental property of AGNs. Traditionally, the BH masses are estimated by assuming that the broad-line clouds are virialized (e.g., Krolik et al. 1991; Wandel et al. 1999; Wu et al. 2004; Greene \& Ho 2005). Another approach to estimate the mass of BH is to use the correlation between BH mass and stellar velocity dispersion by direct or indirect measurement of stellar velocity dispersion (Gebhardt et al. 2000; Ferrarese \& Merritt 2000; Wu \& Han 2001), which is always used to estimate the BH masses in BL Lacs (Woo \& Urry 2002; Wu et al. 2002). The BH masses can also be derived using the relation between the BH mass and host-galaxy bulge luminosity (e.g., Laor 2001). There are 13 GeV-FSRQs in our sample for which the BH masses are obtained from literature, as listed in Table 2; the BH masses of 11 sources are estimated with the traditional virial method. For the other two soueces, one is derived with the relation of BH mass and bulge luminosity and the other one is estimated with the variability timescale.

\section{SED Modeling and Results}
As mentioned in Section 2, there are significant blue bump components at $\sim10^{15}$ Hz in the observed SEDs of GeV-FSRQs in our sample, which are due to the thermal emission of accretion disk. Therefore, our model includes both the jet emission and accretion disk emission and the details of the model are described as follows.

\subsection{Accretion Disk Radiation}

The standard accretion disk spectrum in Davis \& Laor (2011) is used to explain the significant bumps at ultraviolet band in the observed SEDs. The parameters include the inside and outside radius ($R_{\rm in}$ and $R_{\rm out}$) of the accretion disk, BH mass, Eddington ratio, and inclination to the line of sight $\theta$. The BH mass is fixed and the fixed values of BH masses for the sources are listed in Table 2\footnote{The BH masses of the five sources for which no BH mass is found in literature are taken as the average of others, i.e., $M_{\rm BH}=1.1\times10^{9}M_{\bigodot}$.}. We take $R_{\rm out}=500R_{\rm s}$ ($R_{\rm s}$ is the Schwarzschild radius) and $\cos \theta=1$, and then vary both $R_{\rm in}$ (from 0.5$R_{\rm s}$ to 4.5$R_{\rm s}$) and the Eddington ratio to model the accretion disk emission at ultraviolet band.

\subsection{Jet Radiation}

The issue that the radiation regions are inside or outside the BLRs is very important to understand the properties of radiation regions, but highly debated. Considering the limits of Klein-Nashina (KN) effect to explain the observation data at TeV band and the absorption of $\gamma$-rays via photon-photon pair production above 10 GeV (Liu \& Bai 2006) of BLR photon field, some authors suggested that the emission regions should be outside the BLR and the photons from the torus dominate the IC process (Sikora et al. 2009; Tavecchio \& Mazin 2009; Tavecchio et al. 2013). However, Poutanen \& Stern (2010) reported that the spectral breaks in the 2$-$10 GeV range can be well reproduced by the absorption of $\gamma$-rays via photon-photon pair production on He~{\scriptsize II} Lyman recombination continuum and lines, implying that the radiation regions of blazars are inside the highest ionization region of BLR. Isler et al. (2013) also reported that the correlation of the increased emission line flux with millimeter core ejections and $\gamma$-ray, optical, and ultraviolet flares implies that the BLR extends beyond the $\gamma$-emitting region during the 2009 and 2010 flares for 3C 454.3. Sometimes the different timescales of variability are used to investigate the place of radiation region (e.g., Sikora et al. 2009). Some authors used two-zone model to explain the observed SED for the above controversies (e.g., Tavecchio et al. 2011). All the observed highest photon energies in our sample are smaller than 10 GeV, except for PKS 0727-11, which is $\sim$20 GeV, and thus we still assume that the radiation regions are inside the BLRs and the external photon field is dominated by BLR, not torus. Therefore, the single-zone synchrotron+IC model is used to explain the jet emission, where the IC process includes both the SSC process and the external Compton scattering of the BLR photons (EC/BLR). Although the KN effect and the absorption of high energy gamma-ray photons by extragalactic background light (Franceschini et al. 2008) are not significant in the model calculation, we still take them into account.

The radiation region is assumed to be a sphere with radius $R$, which is obtained with $R=\delta c\Delta t/(1+z)$, where $\delta$ is the beaming factor, $\Delta t$ is the variability timescale and taken with $\Delta t=12$ hr following our previous works (Zhang et al. 2012, 2014). The electron distribution is taken as a broken power law, which is characterized by an electron density parameter ($N_0$, in units of cm$^{-3}$), a break energy $\gamma_{\rm b}$ and indices ($p_1$ and $p_2$) in the range of  $\gamma_{\rm e}\sim[\gamma_{\min}, \gamma_{\max}]$. The indices of $p_1$ and $p_2$ are derived from the spectral indices of the observed SEDs as reported by Zhang et al. (2012). $\gamma_{\max}$ can be only poorly constrained by the last data point in the SEDs and is usually dozens of times of $\gamma_{\rm b}$, but it does not significantly affect our results. The energy density of BLR in the jet is boosted by a factor of $\Gamma^2$ ($\Gamma$ is the bulk Lorentz factor) and a corrected factor of 17/12 for $U_{\rm BLR}$ proposed by Ghisellini \& Madau (1996) is also considered because the BLR is not uniform and has an angle to the disk. So the derived energy density of the BLR photon fields in the comoving frame is $U^{'}_{\rm BLR}=\frac{17}{12}\Gamma^2U_{\rm BLR}$. The values of $\frac{17}{12}U_{\rm BLR}$ are listed in Table 1. For the sources whose $U_{\rm BLR}$ are not available, the average of $\frac{17}{12}U_{\rm BLR}=2.73\times10^{-2}$ erg cm$^{-3}$ is taken. Since for blazars we are likely looking at the jet within the $1/\Gamma$ cone, and that the probability is the highest at the rim of the cone, we take $\delta=\Gamma$ in all the calculations, i.e., the viewing angle is equal to the opening angle of the jet.

The free parameter set of our SED modeling is $\left\{B,\delta, N_{0}, \gamma_{\rm b}, \gamma_{\rm min}\right\}$, where $B$ (in units of G) is the magnetic field strength of the radiation region. Following our previous works (Zhang et al. 2012, 2014), the $\chi^2$ minimization technique is also used to perform the SED fits. Since no errors are available for some data points in the selected SEDs, we estimate their errors with the average relative errors of the data points whose errors are available in order to obtain the goodness of the SED fit with the $\chi^2$ minimization technique. The average relative errors of the data for the bumps are derived separately (with a separation at $\nu=10^{20}$ Hz; e.g., Zhang et al. 2012, 2014) because the relative errors of the data for the synchrotron bump and SSC bump are usually much smaller than those for the EC bump. For 3C 454.3, PKS 1510$-$089, and PKS 2123$-$463, no error of observation data below $\nu=10^{20}$ Hz is available, and we take 10\% of the observation flux as the errors (Zhang et al. 2012; 2014). For the details of this technique and fitting strategies please refer to Zhang et al. (2012, 2014).

\subsection{Modeling Results}

\begin{figure*}
\includegraphics[angle=0,scale=0.51]{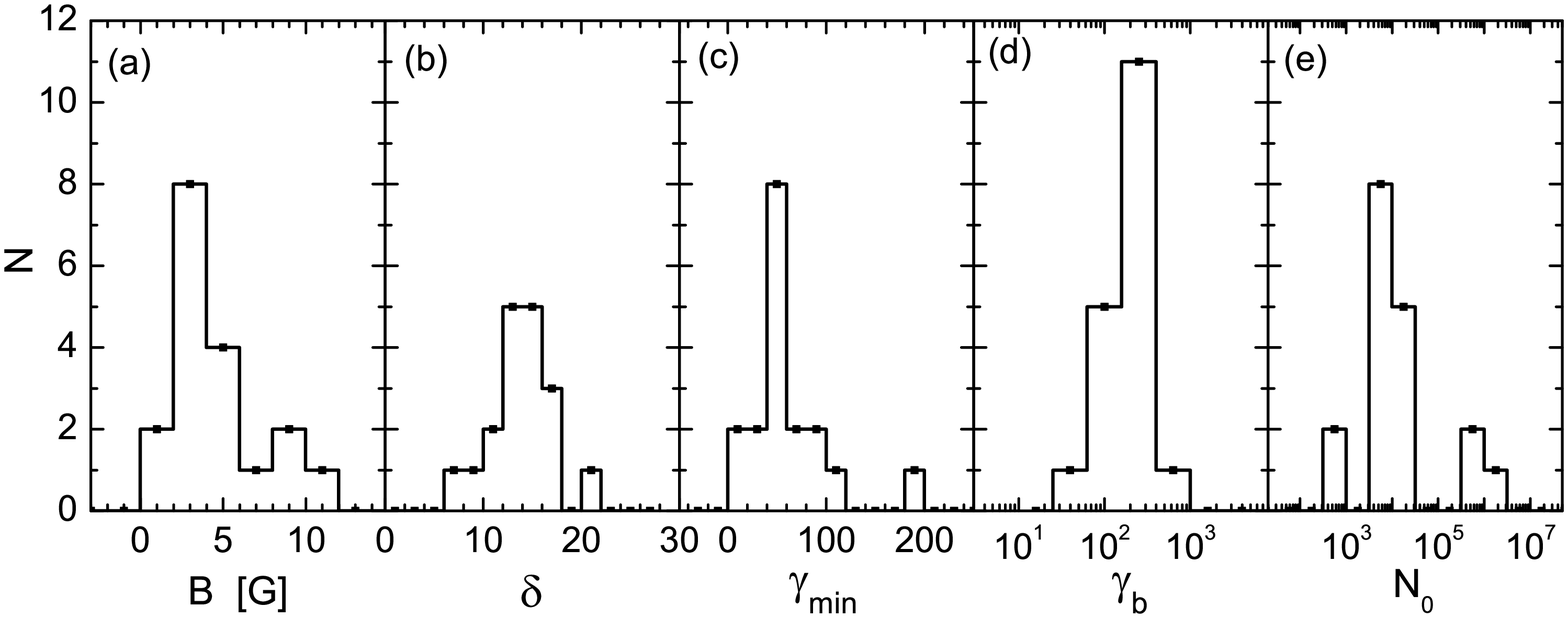}
\caption{Distributions of the magnetic field strength ($B$), the beaming factor ($\delta$), the minimum Lorentz factor of electrons ($\gamma_{\rm min}$),  the break Lorentz factor of electrons ($\gamma_{\rm b}$), and the electron density paramete ($N_{0}$) for the 18 GeV-FSRQs in our sample.}\label{jetpara}
\end{figure*}

Note that we only re-fit the jet emission for PKS 0528+134 and PMN 2345$-$1555 in this paper since the radiation of their accretion disks is taken as jet emission in Zhang et al. (2014), but the jet parameters for the other five sources (3C 273, PKS 0420$-$01, PKS 0727$-$11, 1Jy 1308+326, PKS 1510$-$089) in Zhang et al. (2014) do not change. The best SED fits are shown in Figure 1 and the derived model parameters in the 1$\sigma$ confidence level are reported in Table 2. The distributions of jet fitting parameters $B$, $\delta$, $N_{0}$, $\gamma_{\rm b}$, and $\gamma_{\min}$ are shown in Figure \ref{jetpara}. $\delta$ ranges from 7.4 to 20.6 with a median of $\sim$14. $B$ is from 1.3 to 11.5 with a median of $\sim$3.7. Most of $\gamma_{\min}$ of the sources cluster within 10$\sim$100 with a median of $\sim$50; however, only one source (PKS 2142$-$758) needs a large $\gamma_{\min}\sim197$ to fit the observed spectrum for the very hard spectrum at X-ray band. The distribution of $\gamma_{\rm b}$ clusters within 100$\sim$400, but the source TXS 2141+175 needs very small $\gamma_{\min}$ and $\gamma_{\rm b}$ to explain the observed spectrum at X-ray and GeV bands. The values of $N_{0}$ mostly cluster in $3\times10^3-3\times10^4$ cm$^{-3}$. From the fitting spectra, we also calculate the luminosities of accretion disk radiation ($L_{\rm disk}$) and jet radiation ($L_{\rm jet}$). The values of $L_{\rm disk}$ and $L_{\rm jet}$ are also reported in Table 2.

\section{Correlations Between Accretion Disk Luminosity and BLR Luminosity}

\begin{figure*}
\includegraphics[angle=0,scale=0.3]{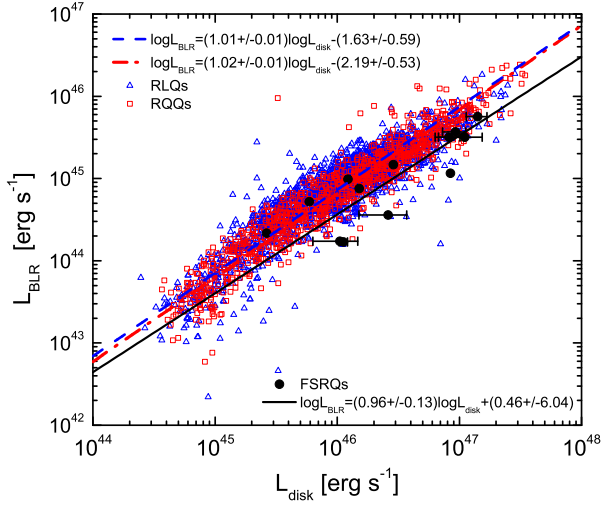}
\caption{$L_{\rm BLR}$ as a function of $L_{\rm disk}$ for the 13 GeV-FSRQs (black solid circles) in our sample. The data of the RLQs (blue opened triangles) and RQQs (red opened squares) are from Shen et al. (2011). The lines are the bisectors of the two ordinary least-squares lines of linear regression fits to the data for GeV-FSRQs (black solid line), RLQs (blue dashed line), and RQQs (red dash-dotted line), respectively.}\label{corr-Ldisk-Lblr}
\end{figure*}

As described in Section 2.2, a sub-sample of BLR luminosity ($L_{\rm BLR}$) for the 13 GeV-FSRQs in our sample is obtained, and the luminosity of accretion disk ($L_{\rm disk}$) can be estimated from fitting the blue bump. $L_{\rm BLR}$ for the 13 GeV-FSRQs as a function of $L_{\rm disk}$ is presented in Figure \ref{corr-Ldisk-Lblr}. $L_{\rm BLR}$ is strongly correlated with $L_{\rm disk}$ for the 13 GeV-FSRQs with a Pearson correlation coefficient of $r=0.85$ and chance probability of $p=2.7\times10^{-4}$. The linear fit in the log scale gives $\log L_{\rm BLR}=(0.46\pm6.04)+(0.96\pm0.13)\log L_{\rm disk}$. Here and hereafter, the bisector of the two ordinary least-squares lines of linear regression fit is used in order to avoid specifying independent and dependent variables (Isobe et al. 1990). With this correlation, one can estimate the accretion disk luminosity using the luminosity of BLR when the nonthermal emission of jet overwhelms the thermal emission of accretion disk.

In order to compare the $L_{\rm BLR}$--$L_{\rm disk}$ relation of GeV-FSRQs with the normal quasars, the sample of the 105,783 quasars from Shen et al. (2011) is compiled, which is taken in the Sloan Digital Sky Survey Data Release 7 quasar catalog. We only select the sources that have $\rm H\beta$ or/and Mg~{\scriptsize II} emission lines\footnote{Compared with the emission line of C~{\scriptsize IV}, the relative fluxes of H$\beta$ and Mg~{\scriptsize II} are more consistent with that reported by Francis et al. (1991), and the redshift range of the selected sample is also more consistent with ours.} with the signal to noise ratio of the lines larger than 10. To include radio properties, we compile two sub-samples with the radio loudness of $R_{\rm RL}=f_{\rm 6~cm}/f_{2500}$, where $f_{\rm 6~cm}$ and $f_{2500}$ are the flux density at rest-frame 6 cm and 2500 ${\rm \AA}$, and then we get 1999 very radio-loud quasars with $R_{\rm RL}>100$ (hereafter RLQs) and 937 radio-quiet quasars with $R_{\rm RL}<10$ (hereafter RQQs). In Shen et al. (2011), the bolometric luminosities ($L_{\rm bol}$) are derived from the continuum luminosities at 5100 ${\rm \AA}$ ($L_{5100}$, $z<0.7$), 3000 ${\rm \AA}$ ($L_{3000}$, $0.7\leq z<1.9$), and 1350 ${\rm \AA}$ ($L_{1350}$, $z\geq 1.9$) with the correlations between $L_{\rm bol}$ and continuum luminosities in Richards et al. (2006). The values of $L_{\rm disk}$ for the selected samples of RQQs and RLQs in Shen et al. (2011) are derived by $\frac{1}{2}L_{\rm bol}$ (Calderone et al. 2013). The luminosities of BLRs for these quasars are also calculated with the fluxes of emission lines ($\rm H\beta$ or/and Mg~{\scriptsize II}) using the method proposed in Celotti et al. (1997). The data of the selected RQQs and RLQs in Shen et al. (2011) are also presented in the $L_{\rm BLR}$--$L_{\rm disk}$ plane, and $L_{\rm BLR}$ are strongly correlated with $L_{\rm disk}$ for the selected RQQs and RLQs, respectively. The linear fits in the log scale give $\log L_{\rm BLR}=-(2.19\pm0.53)+(1.02\pm0.01)\log L_{\rm disk}$ for RQQs and $\log L_{\rm BLR}=-(1.63\pm0.59)+(1.01\pm0.01)\log L_{\rm disk}$ for RLQs, respectively.

It can be found that the slope of the $L_{\rm BLR}$--$L_{\rm disk}$ relation for GeV-FSRQs is consistent with that of both RQQs and RLQs, but the intercepts are slightly different among the three samples. We test whether the $L_{\rm BLR}$--$L_{\rm disk}$ relations between GeV-FSRQs with RQQs and RLQs show any statistical difference with the two-dimensional Kolmogorov$-$Smirnov test (K--S test), which yields a chance probability of $p_{\rm KS}$. A two-dimensional K$-$S test probability larger than 0.2 (0.1 for one-dimensional) would strongly suggest no statistical difference between two samples, and a K$-$S test probability smaller than 10$^{-4}$ would strongly suggest there is statistical difference between two samples. All the values of $p_{\rm KS}$ derived with the two-dimensional (also the following one-dimensional) K$-$S tests between two samples are presented in Table 3. The results of K$-$S tests show that the 13 GeV-FSRQs in the $L_{\rm BLR}$--$L_{\rm disk}$ plane are not statistically distinct from the RQQs ($p_{\rm KS}=0.29$), but they are marginally distinct from the RLQs ($p_{\rm KS}=0.045$).

\begin{figure*}
\includegraphics[angle=0,scale=0.54]{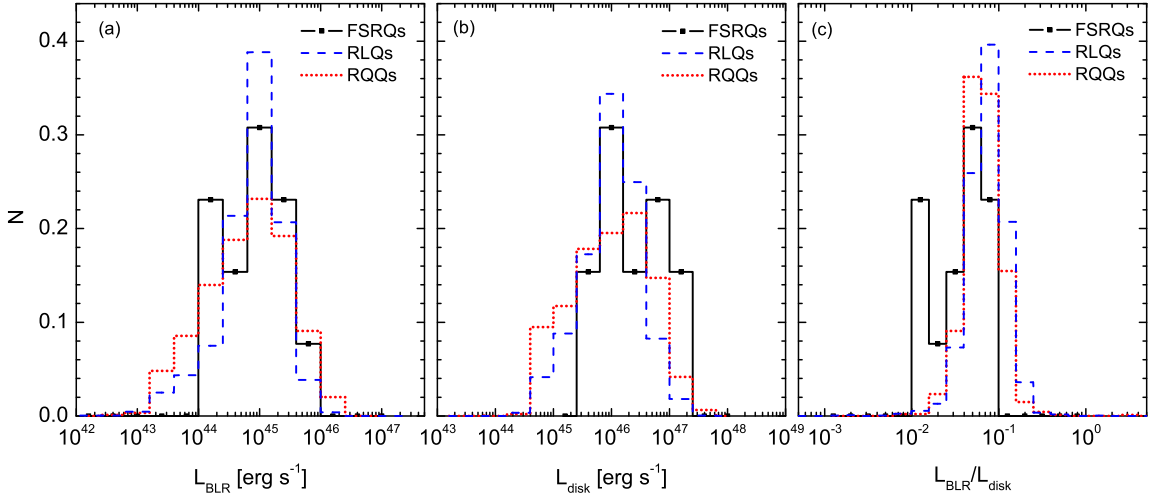}
\caption{Normalized distributions of the BLR luminosity ($L_{\rm BLR}$), the disk luminosity ($L_{\rm disk}$), and the BLR covering factor ($L_{\rm BLR}/L_{\rm disk}$) for the 13 GeV-FSRQs (black solid lines) in our sample. The data of the RLQs (blue dashed lines) and RQQs (red dotted lines) are from Shen et al. (2011).}\label{dis-Ldisk-Lblr}
\end{figure*}

To further investigate this issue, the distributions of $L_{\rm BLR}$ and $L_{\rm disk}$ \footnote{The values of $L_{\rm disk}$ only for the GeV-FSRQs in which $L_{\rm BLR}$ is available are taken into account in this section, hence there are 13 sources, not 18.} as well as $L_{\rm BLR}/L_{\rm disk}$ for the 13 GeV-FSRQs and the selected samples of RQQs and RLQs are given in Figure \ref{dis-Ldisk-Lblr}. Considering the much different numbers of sources in the different samples, the distributions are normalized (i.e., the total number of the sources in the distribution has been set to unity) as shown in Figure \ref{dis-Ldisk-Lblr}. The mean of $L_{\rm BLR}/L_{\rm disk}$ for the 13 GeV-FSRQs is 0.043, different from 0.082 and 0.076 for RLQs and RQQs, respectively. The one-dimensional K--S tests indicate that the distribution of $L_{\rm BLR}$ for the 13 GeV-FSRQs is analogous to that of both RQQs and RLQs, but the distribution of $L_{\rm disk}$ for the 13 GeV-FSRQs is marginally different from that of both RQQs and RLQs in Shen et al. (2011), and thus the distribution of covering factor ($L_{\rm BLR}/L_{\rm disk}$) for the 13 GeV-FSRQs is different from that of the large quasar samples, either RQQs or RLQs. This also results in the similar slopes but the slightly different intercepts among the three samples in the $L_{\rm BLR}$--$L_{\rm disk}$ plane.

\section{Correlations Between Accretion Disk Luminosity and Jet Power}

\begin{figure*}
\includegraphics[angle=0,scale=0.52]{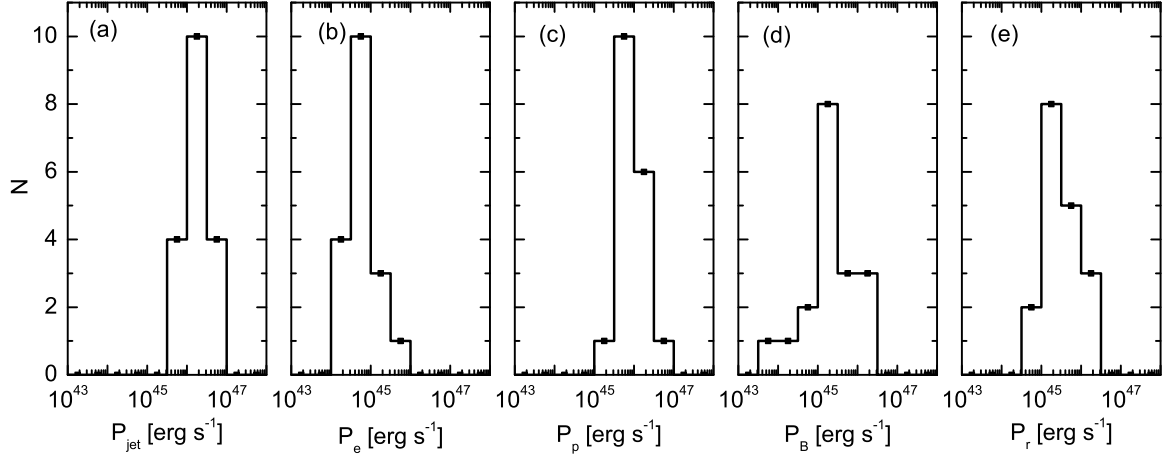}
\caption{Distributions of the jet power ($P_{\rm jet}$) and the powers associated with relativistic electrons ($P_{\rm e}$), cold protons ($P_{\rm p}$), Poynting flux ($P_{\rm B}$), radiation component ($P_{\rm r}$) for the 18 GeV-FSRQs in our sample.}\label{Pjet}
\end{figure*}

The jet power is estimated with the SED fitting parameters by assuming that they are carried by relativistic electrons, cold protons, magnetic fields, and radiation (e.g., Ghisellini et al. 2010; Zhang et al. 2012, 2013, 2014), i.e., $P_{\rm jet}=\sum_i P_{\rm i}$, where $P_{\rm i}=2\pi R^2 \Gamma^2 c U^{'}_{\rm i}$ (the factor of 2 accounts for twin jets) and $U^{'}_{i} (i={\rm e,\ p,}\ B, \rm r)$ are the powers and the energy densities associated with the emitting electrons, cold protons, magnetic fields, and radiation (e.g., Zhang et al. 2013, 2014), where the proton-electron pair assumption is used. The distributions of jet power and each ingredient power for the 18 GeV-FSRQs are shown in Figure \ref{Pjet}. $P_{\rm jet}$ ranges within $5\times10^{45}-8\times10^{46}$ erg s$^{-1}$. The distribution medians of jet ingredient powers are $5.85\times10^{44}$ erg s$^{-1}$, $9.14\times10^{45}$ erg s$^{-1}$, $2.31\times10^{45}$ erg s$^{-1}$, and $3.08\times10^{45}$ erg s$^{-1}$ for $P_{\rm e}$, $P_{\rm p}$, $P_{B}$, and $P_{\rm r}$, respectively. The means of the jet radiation efficiency ($\varepsilon_{\rm r}=P_{\rm r}/P_{\rm jet}$) and the jet magnetization parameter ($\sigma=P_{\rm B}/(P_{\rm p}+P_{\rm e}+P_{\rm r})$, Zhang et al. 2013, 2014) for the 18 GeV-FSRQs in our sample are 0.22 and 0.29, respectively, which is consistent with the result of our previous works that the FSRQ jets are highly magnetized with high radiation efficiency.

\begin{figure*}
\includegraphics[angle=0,scale=0.2]{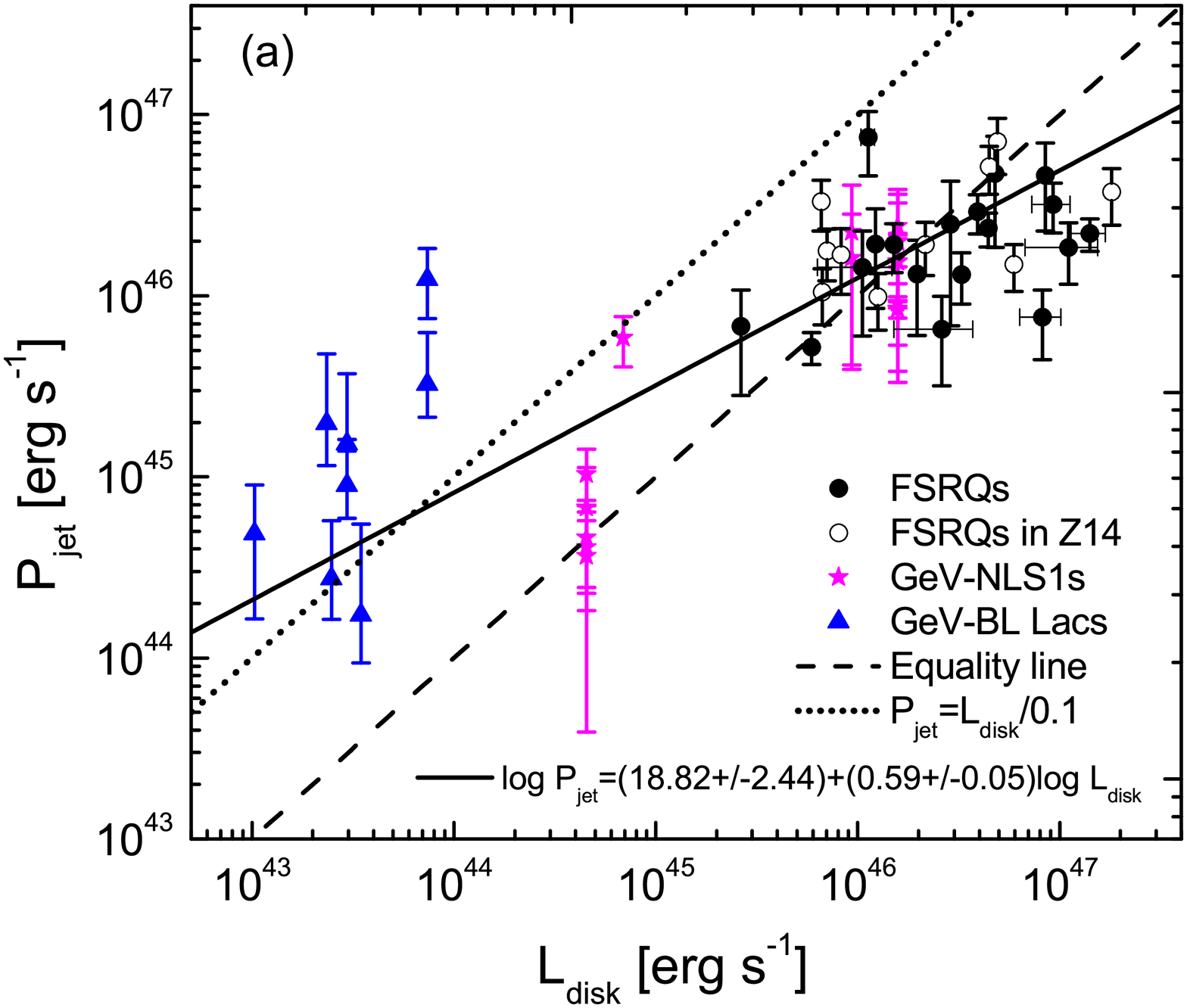}
\includegraphics[angle=0,scale=0.2]{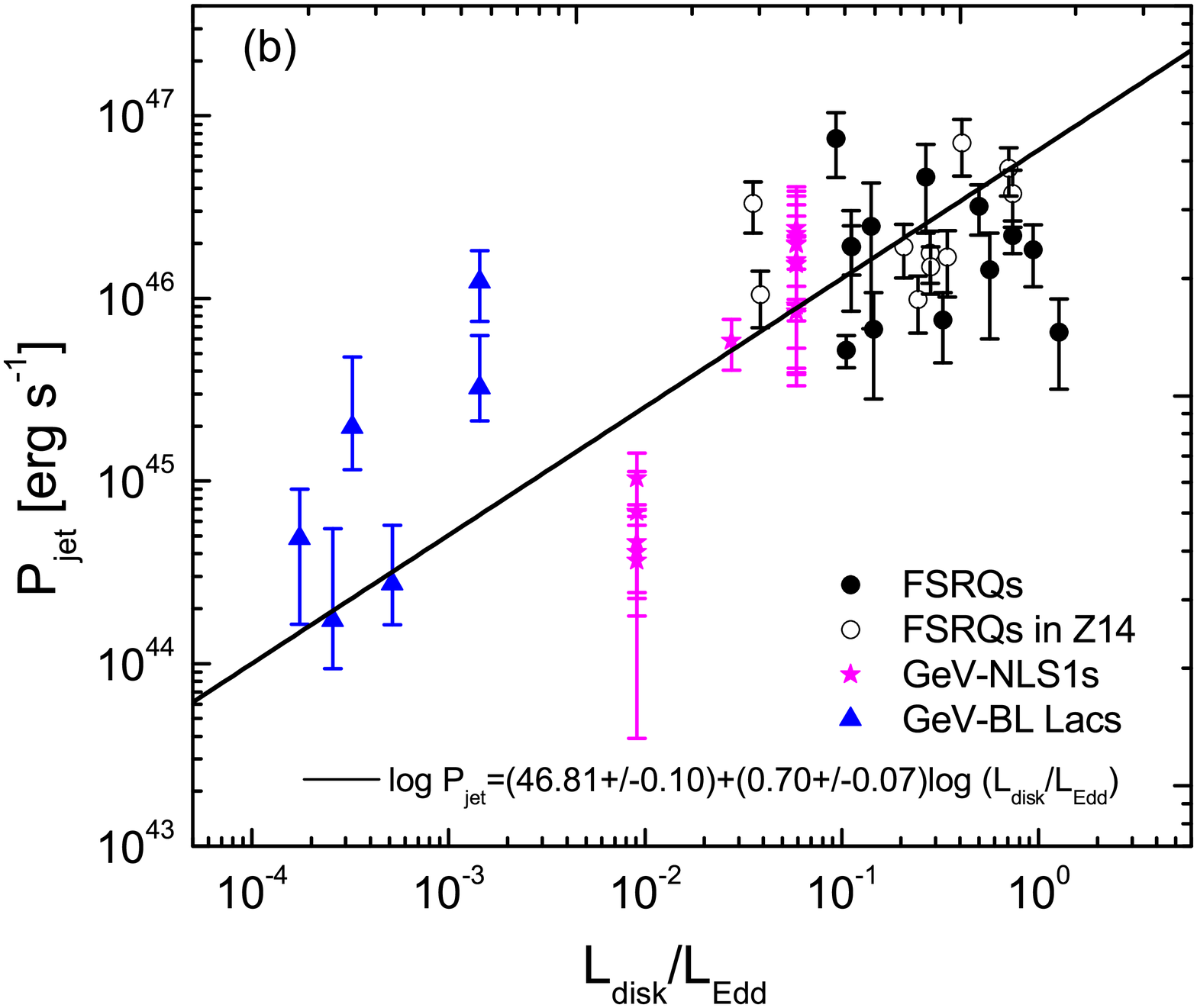}
\includegraphics[angle=0,scale=0.2]{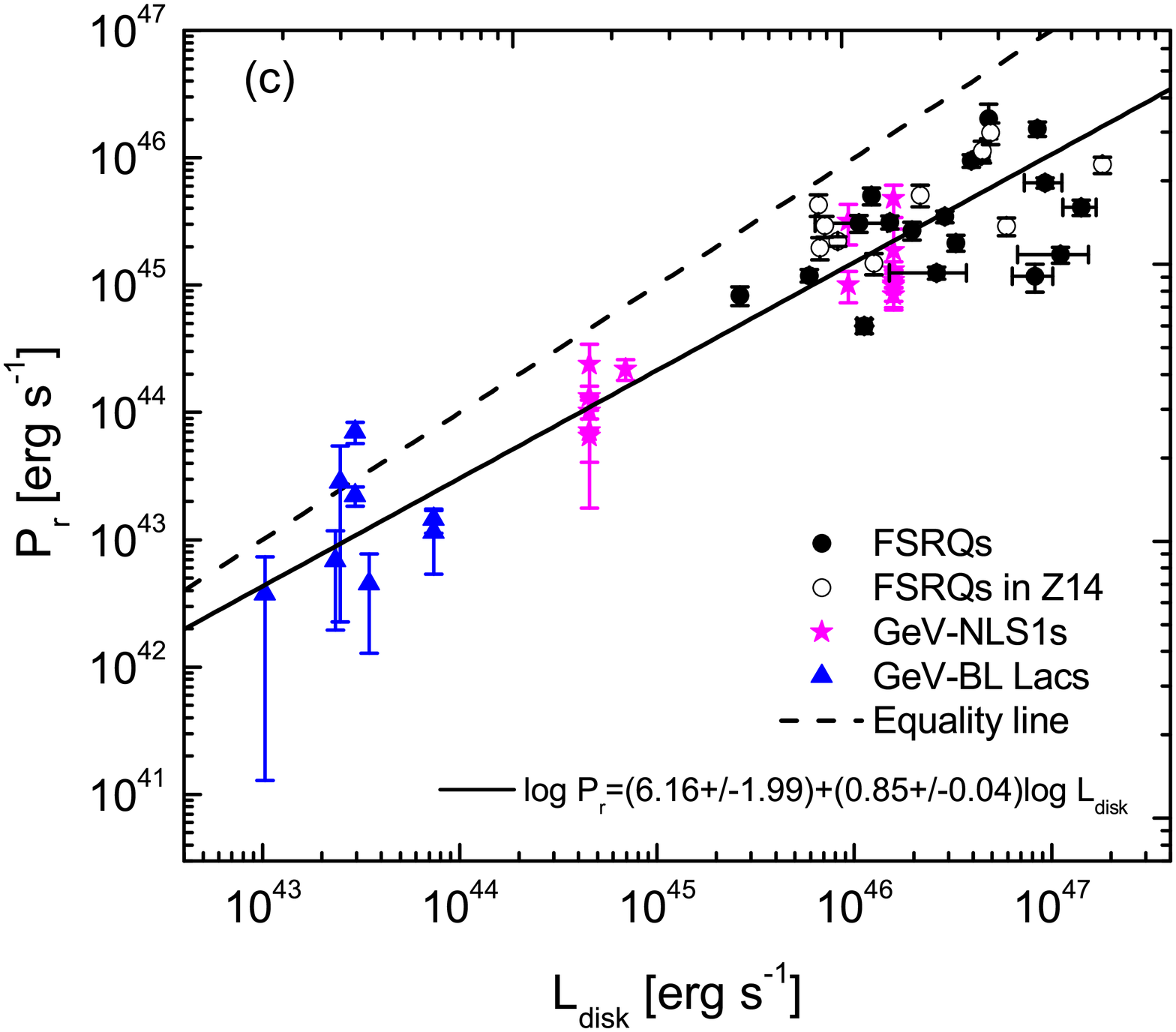}
\caption{$P_{\rm jet}$ as functions of $L_{\rm disk}$ (Panel (a)) and $L_{\rm disk}/L_{\rm Edd}$ (Panel (b)), and $P_{\rm r}$ as a function of $L_{\rm disk}$ (Panel (c)) for the GeV-FSRQs (black solid circles) in our sample. The data of the 10 GeV FSRQs (black opened circles) in Zhang et al. (2014), the BL Lacs (blue triangles) with available $L_{\rm BLR}$ in Zhang et al. (2012), and the GeV NLS1 galaxies (magenta stars) from Sun et al. (2015) are also presented for comparison. The solid lines are the bisectors of the two ordinary least-squares lines of linear regression fits for all the sources in the figures, including FSRQs, NLS1 galaxies, and BL Lacs.}\label{Pjet-Pr-Ldisk}
\end{figure*}

The jet power ($P_{\rm jet}$) as functions of the accretion disk luminosity ($L_{\rm disk}$) and the Eddington ratio ($L_{\rm disk}/L_{\rm Edd}$), and the jet radiation power ($P_{\rm r}$) as a function of $L_{\rm disk}$ are given in Figure \ref{Pjet-Pr-Ldisk}. There are 10 sources  in Zhang et al. (2014, hereafter FSRQs in Z14) that are not included in the selected sample due to lack of significant blue bump in their SEDs, for which the data of $L_{\rm BLR}$ are available in literature, as listed in Table 4. There are also some GeV BL Lacs \footnote{Including six sources, eight SEDs (two states with well constraints on $\gamma_{\min}$ for BL Lacertae and W Com), they are Mkn 421, Mkn 501, 1ES 1218+304, W Com, PKS 2005$-$489, and BL Lacertae as given in Table 4 and Zhang et al. (2012).} in Zhang et al. (2012) for which $L_{\rm BLR}$ is available in literature as listed in Table 4. Thus the data of the 10 FSRQs in Z14 and some BL Lacs in Zhang et al. (2012) are also shown in Figure \ref{Pjet-Pr-Ldisk}, and $L_{\rm disk}$ is derived from $L_{\rm BLR}$ with the correlation between $L_{\rm disk}$ and $L_{\rm BLR}$ in Section 4 (Figure \ref{corr-Ldisk-Lblr}). For comparison, the data of a GeV NLS1 galaxy sample taken from Sun et al. (2015) are also given in Figure \ref{Pjet-Pr-Ldisk}. Note that the values of $P_{\rm jet}$ and $P_{\rm r}$ for the 10 FSRQs in Z14, BL Lacs in Zhang et al. (2012), and GeV NLS1 galaxies in Sun et al. (2015) have been multiplied by 2 to account for twin jets.

In the $L_{\rm disk}$--$P_{\rm jet}$ plane, one can find that $L_{\rm disk}$ is higher than $P_{\rm jet}$ for most of GeV FSRQs, but not so for BL Lacs in which the derived $L_{\rm disk}$ is lower than $P_{\rm jet}$. The data of GeV NLS1 galaxies are in a broad range and are intermediate between GeV FSRQs and BL Lacs, as reported by Sun et al. (2015); hence the GeV NLS1 galaxies are indeed a bridge between the FSRQs and BL Lacs. Assuming $L_{\rm disk}=0.1\dot{M}c^2$, where $\dot{M}$ is the accretion rate, the available accretion power of BH can be estimated by $\dot{M}c^2=L_{\rm disk}/0.1$. As shown in Figure \ref{Pjet-Pr-Ldisk}(a), $P_{\rm jet}$ of all the FSRQs and GeV NLS1 galaxies is lower than $\dot{M}c^2$, but not so for most of BL Lacs. This may be due to that it is not a standard thin disk in the center of BL Lacs, but a radiatively inefficient disk (e.g., Zhang et al. 2014; Sun et al. 2015); in this case, $L_{\rm disk}$ is lower than $0.1\dot{M}c^2$ (Yuan \& Narayan 2014), i.e., $\dot{M}c^2>L_{\rm disk}/0.1$ for BL Lacs. Therefore, intrinsically $P_{\rm jet}$ of BL Lacs may be also lower than $\dot{M}c^2$. No correlation between $L_{\rm disk}$ and $P_{\rm jet}$ is observed for the 18 GeV-FSRQs in our sample, even taking the 10 FSRQs in Z14 into account. However, if the data of BL Lacs and GeV NLS1 galaxies are taken into account, $P_{\rm jet}$ would be strongly correlated with $L_{\rm disk}$ with the Pearson correlation coefficient of $r=0.82$ and chance probability of $p=1.5\times10^{-13}$. The linear fit in the log scale gives $\log P_{\rm jet}=(18.82\pm2.44)+(0.59\pm0.05)\log L_{\rm disk}$. Moreover, $P_{\rm jet}$ is also strongly correlated with $L_{\rm disk}/L_{\rm Edd}$ for FSRQs, NLS1 galaxies, and BL Lacs as shown in Figure \ref{Pjet-Pr-Ldisk}(b); the Pearson correlation coefficient of $r=0.75$ and chance probability of $p=4.7\times10^{-9}$.

In the $L_{\rm disk}$--$P_{\rm r}$ plane, $P_{\rm r}$ for all the sources is lower than their $L_{\rm disk}$, except for one BL Lac. A tentative correlation between the two parameters seems to be observed for the 18 GeV-FSRQs in our sample with the Pearson correlation coefficient of $r=0.48$ and chance probability of $p=0.042$. If taking the 10 FSRQs in Z14 into account, the tentative correlation becomes slightly stronger with $r=0.50$ and $p=0.007$. It is interesting to find that BL Lacs, GeV NLS1 galaxies, and FSRQs form a sequence in the $L_{\rm disk}$--$P_{\rm r}$ plane; the linear fit in the log scale gives $\log P_{\rm r}=(6.16\pm1.99)+(0.85\pm0.04)\log L_{\rm disk}$ with a Pearson correlation coefficient of $r=0.93$ and chance probability of $p\sim0$.

These results may indicate that the total accretion power of BH is sufficient to launch the jets in GeV-FSRQs, at least for the jet radiation in low state when the accretion disk radiation is not overwhelmed by the jet radiation; however the lack of strong correlation between $L_{\rm disk}$ and $P_{\rm jet}$ may imply that their energy sources are not exactly the same for GeV-FSRQs. But the strong correlations between $P_{\rm jet}$ with $L_{\rm disk}$ and $L_{\rm disk}/L_{\rm Edd}$ for FSRQs, GeV NLS1 galaxies, and BL Lacs may imply that the Eddington ratio may be the fundamental in the unified framework among different types of GeV AGNs.

Note that above results are based on the traditional assumption of one cold proton for one relativistic electron in the calculations of jet powers. There is no way to directly measure the ratio of protons to electrons in the jets. The proton-electron pair assumption is widely adopted in the calculations of jet powers in blazars (e.g., Ghisellini et al. 2010; Zhang et al. 2012, 2014; Chen et al. 2012). We check another extreme case that the jet power is carried by positron-electron pairs, magnetic field, and radiation, but no protons. In this scenario, the jet powers are still uncorrelated with $L_{\rm disk}$ for the 28 FSRQs, and $P_{\rm jet}$ for the FSRQs and GeV NLS1 galaxies are more lower than the accretion powers of BH, even for half of BL Lacs.

\section{Relations Among Jet Power, BLR Luminosity, and Accretion Disk Luminosity}

\begin{figure*}
\includegraphics[angle=0,scale=0.3]{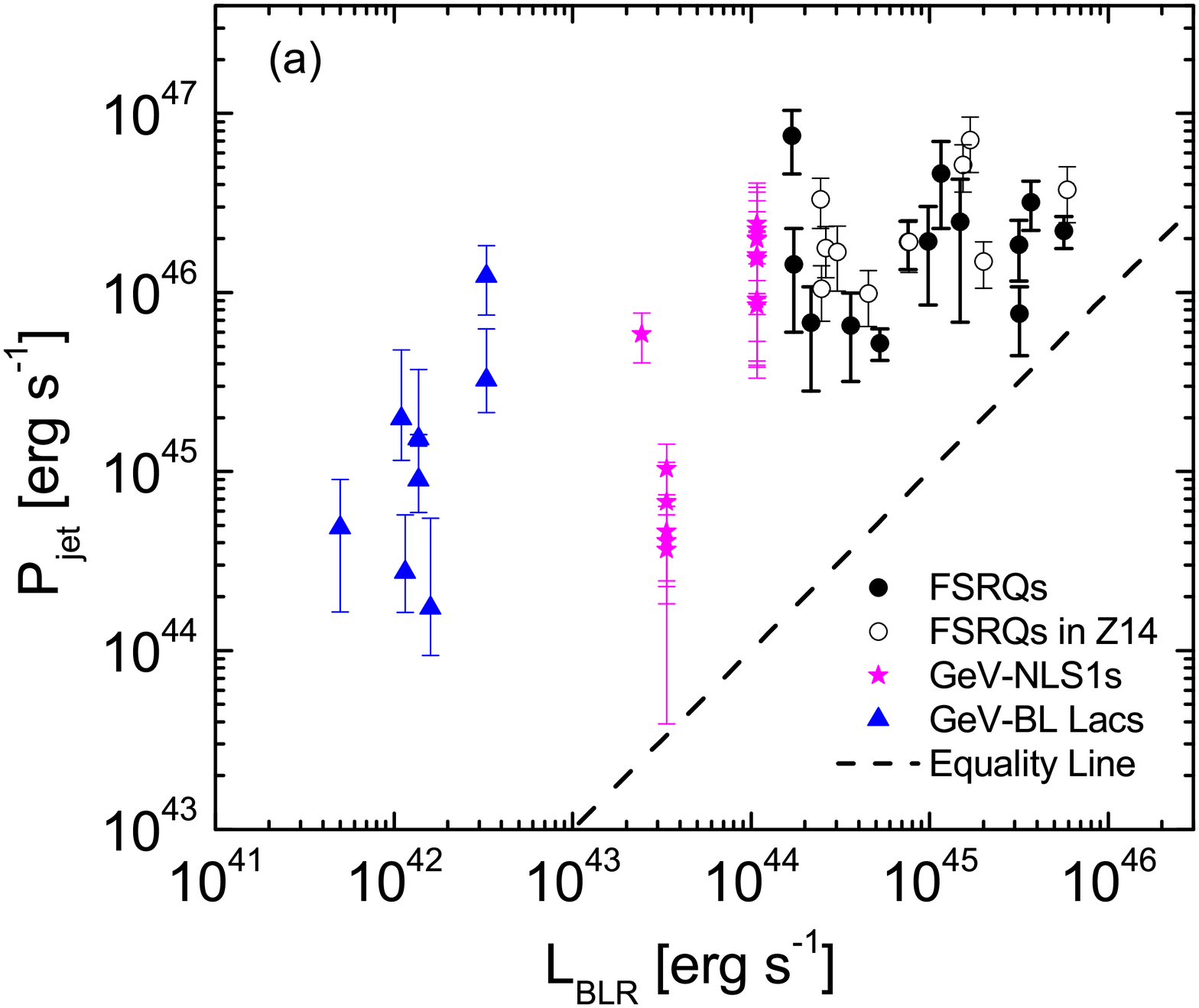}
\includegraphics[angle=0,scale=0.3]{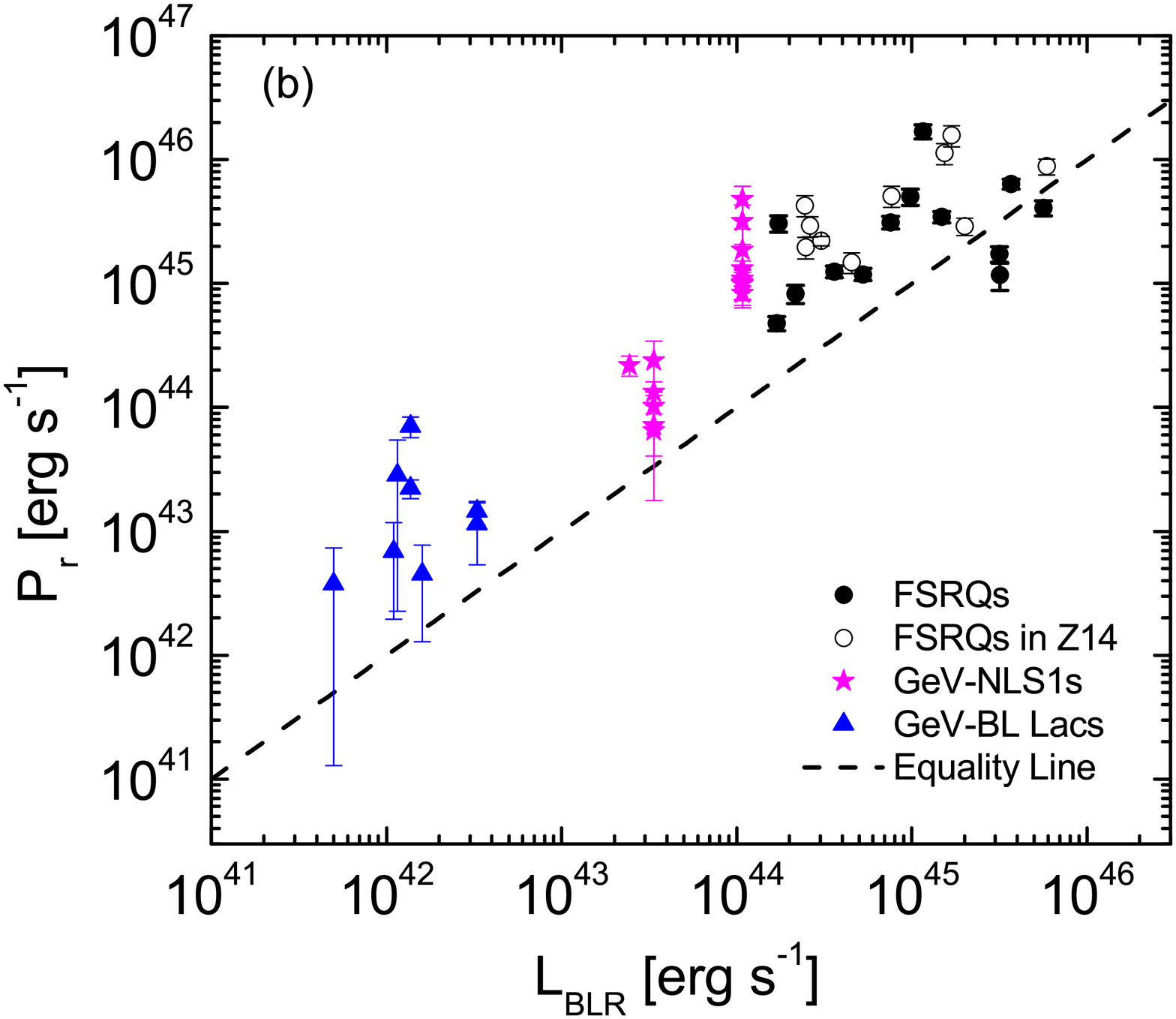}
\caption{$P_{\rm jet}$ and $P_{\rm r}$ as a function of $L_{\rm BLR}$ for the 13 GeV-FSRQs (solid circles) in our sample. The data of the 10 GeV FSRQs (black opened circles) in Zhang et al. (2014), the BL Lacs (blue triangles) with available $L_{\rm BLR}$ in Zhang et al. (2012), and the GeV NLS1 galaxies (magenta stars) from Sun et al. (2015) are also presented for comparison.}\label{Pjet-Pr-Lblr}
\end{figure*}

In order to further study the relations among the jet, accretion disk, and BLR, $P_{\rm jet}$ and $P_{\rm r}$ as a function of $L_{\rm BLR}$ for the 13 GeV-FSRQs in our sample are shown in Figure \ref{Pjet-Pr-Lblr}. The data of the 10 FSRQs in Z14, some BL Lacs in Zhang et al. (2012), and the GeV NLS1 galaxies in Sun et al. (2015) are also presented in Figure \ref{Pjet-Pr-Lblr}. No correlation for FSRQs in the $P_{\rm jet}$--$L_{\rm BLR}$ plane is found. But there is a weak tentative correlation between $P_{\rm r}$ and $L_{\rm BLR}$ for the 23 GeV FSRQs with the Pearson correlation coefficient of $r=0.48$ and chance probability of $p=0.02$. This tentative correlation may be due to that the emission of GeV FSRQs is dominated by EC process, for which the photon field is from BLR. It is similar to the $L_{\rm disk}$--$P_{\rm r}$ plane, BL Lacs, GeV NLS1 galaxies, and FSRQs also form a sequence in the $L_{\rm BLR}$--$P_{\rm r}$ plane.

\begin{figure*}
\includegraphics[angle=0,scale=0.3]{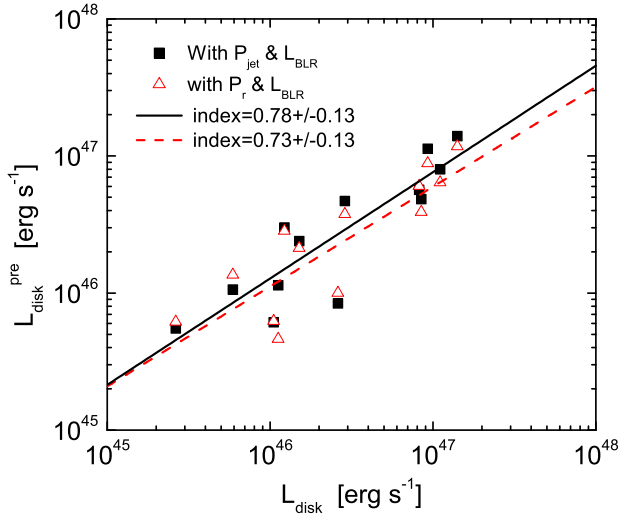}
\caption{The relations of the observed accretion disk luminosity ($L_{\rm disk}$) and the predicted luminosity ($L^{\rm pre}_{\rm disk}$) that are derived with the functions of the three element linear regression of $\log L^{\rm pre_1}_{\rm disk}=-9.7+0.85\log L_{\rm BLR}+0.39\log P_{\rm jet}$ and $\log L^{\rm pre_2}_{\rm disk}=2.3+0.83\log L_{\rm BLR}+0.15\log P_{\rm r}$, respectively. The black solid line and the red dashed line are the best fits for $L_{\rm disk}$ with $L^{\rm pre_1}_{\rm disk}$ and $L^{\rm pre_2}_{\rm disk}$, respectively. }\label{Ldisk-Lblr-Pjet}
\end{figure*}

Then we make multiple linear regressions of the 13 GeV-FSRQs for $L_{\rm disk}$, $L_{\rm BLR}$, $P_{\rm jet}$ and for $L_{\rm disk}$, $L_{\rm BLR}$, $P_{\rm r}$, respectively. The multiple linear regressions yield $\log L_{\rm disk}=(-9.7\pm12.3)+(0.85\pm0.16)\log L_{\rm BLR}+(0.39\pm0.24)\log P_{\rm jet}$ with R-Square of 0.73 and $\log L_{\rm disk}=(2.3\pm10.5)+(0.83\pm0.20)\log L_{\rm BLR}+(0.15\pm0.25)\log P_{\rm r}$ with R-Square of 0.67, respectively. The predicted disk luminosity of $L_{\rm disk}^{\rm pre}$, derived with the above fitting functions of the multiple linear regressions, as a function of the observed accretion disk luminosity ($L_{\rm disk}$) is shown in Figure \ref{Ldisk-Lblr-Pjet}. The best linear fits in the log scale give $L_{\rm disk}^{\rm pre} \propto L_{\rm disk}^{0.78\pm0.13}$ with R-Square of 0.75 and $L_{\rm disk}^{\rm pre} \propto L_{\rm disk}^{0.73\pm0.13}$ with R-Square of 0.70 for using $P_{\rm jet}$ and $P_{\rm r}$, respectively. These results further confirm that $L_{\rm disk}$ for the 13 GeV-FSRQs is tight correlated with $L_{\rm BLR}$, but with neither $P_{\rm jet}$ nor $P_{\rm r}$. This indicates that $L_{\rm disk}$ has the same energy origin as $L_{\rm BLR}$, but not $P_{\rm jet}$; $L_{\rm disk}$ is all from accretion process, but $P_{\rm jet}$ may be dominated by extracting the rotational energy of BH. Hence the dominant mechanism of jet launching in these GeV-FSRQs may be BZ process as discussed in our previous works (Zhang et al. 2014; Sun et al. 2015).

\section{Discussion}

\subsection{Anisotropic Radiation of Accretion Disk}

\begin{figure*}
\includegraphics[angle=0,scale=0.5]{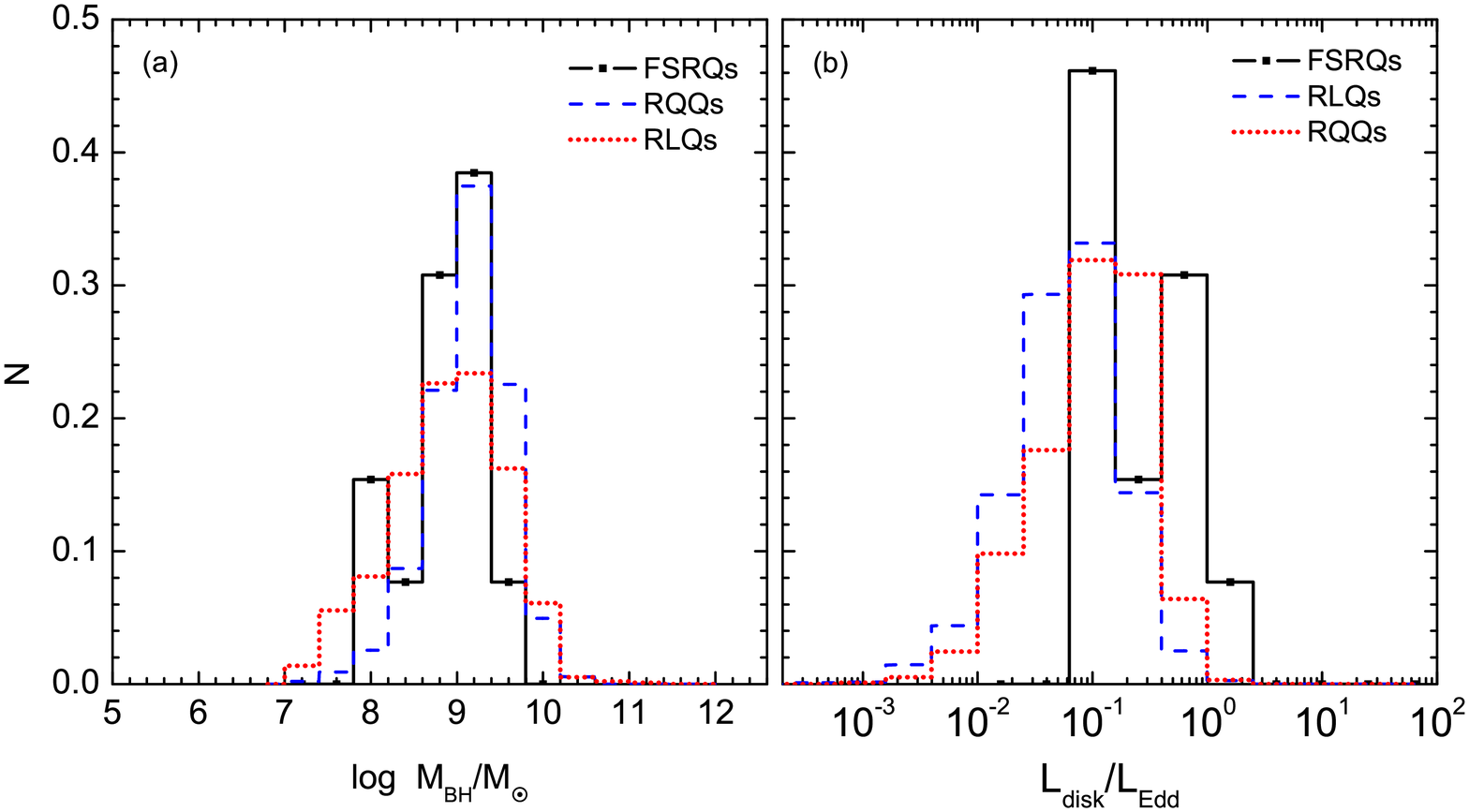}
\caption{Normalized distributions of BH masses and $L_{\rm disk}/L_{\rm Edd}$ for the 13 GeV-FSRQs (black solid lines) in our sample. The data of the RLQs (blue dashed lines) and RQQs (red dotted lines) are from Shen et al. (2011).}\label{BHmass-LEdd}
\end{figure*}

As described in Section 4, the different covering factor $L_{\rm BLR}/L_{\rm disk}$ of BLR for the 13 GeV-FSRQs with the selected RQQs and RLQs should be due to the different distributions of their $L_{\rm disk}$. In order to further compare the states of accretion disk among the three samples, the normalized distributions of BH masses ($M_{\rm BH}$) and Eddington ratio ($L_{\rm disk}/L_{\rm Edd}$) are also shown in Figure \ref{BHmass-LEdd}. The results of K$-$S tests declare that the distribution of $M_{\rm BH}$ for the 13 GeV-FSRQs is similar to that of RLQs and RQQs, but the distribution of $L_{\rm disk}/L_{\rm Edd}$ for the 13 GeV-FSRQs is different from both RLQs and RQQs. Therefore, the main reason for the different $L_{\rm BLR}/L_{\rm disk}$ and $L_{\rm disk}/L_{\rm Edd}$ of the 13 GeV-FSRQs with RLQs and RQQs is the different distributions of their $L_{\rm disk}$.

Note that the GeV-FSRQs in our sample belong to blazars, in which the relativistic jets (also the normal of the accretion disk) are thought to be close to the line of sight. With the same intrinsic luminosity, one may observe the higher flux of accretion disk for the GeV-FSRQs than other sources if the radiation of the accretion disk is anisotropic (Liu \& Zhang 2011). Assuming that the emitting flux of accretion disk depends on the inclination angle, i.e., $F=F_0\cos\theta$, where $\theta$ is the angle between the line of sight and the normal of the accretion disk, the inferred intrinsic luminosity of accretion disk should be $L^{\theta}_{\rm disk}=\frac{L_{\rm disk}}{2\cos\theta}$ (see details in the Appendix), i.e., $L^{\theta}_{\rm disk}=L_{\rm disk}/2$ for the GeV-FSRQs. Assuming that the inclination of accretion disk for RLQs and RQQs is the average value of $\sim 60^{\circ}$, the derived isotropic luminosity of accretion disk is just the inferred intrinsic luminosity of accretion disk for RLQs and RQQs, i.e., $L^{\theta}_{\rm disk}=L_{\rm disk}$. One would observe the similar distributions of BLR covering factor ($L_{\rm BLR}/L^{\theta}_{\rm disk}$) for the GeV-FSRQs with RLQs and RQQs if considering the inclination effect and anisotropic radiation of accretion disk; the BLR covering factor of the GeV-FSRQs would be close to 0.1 with a mean of 0.086, similar to RLQs and RQQs in Shen et al. (2011).

\begin{figure*}
\includegraphics[angle=0,scale=0.3]{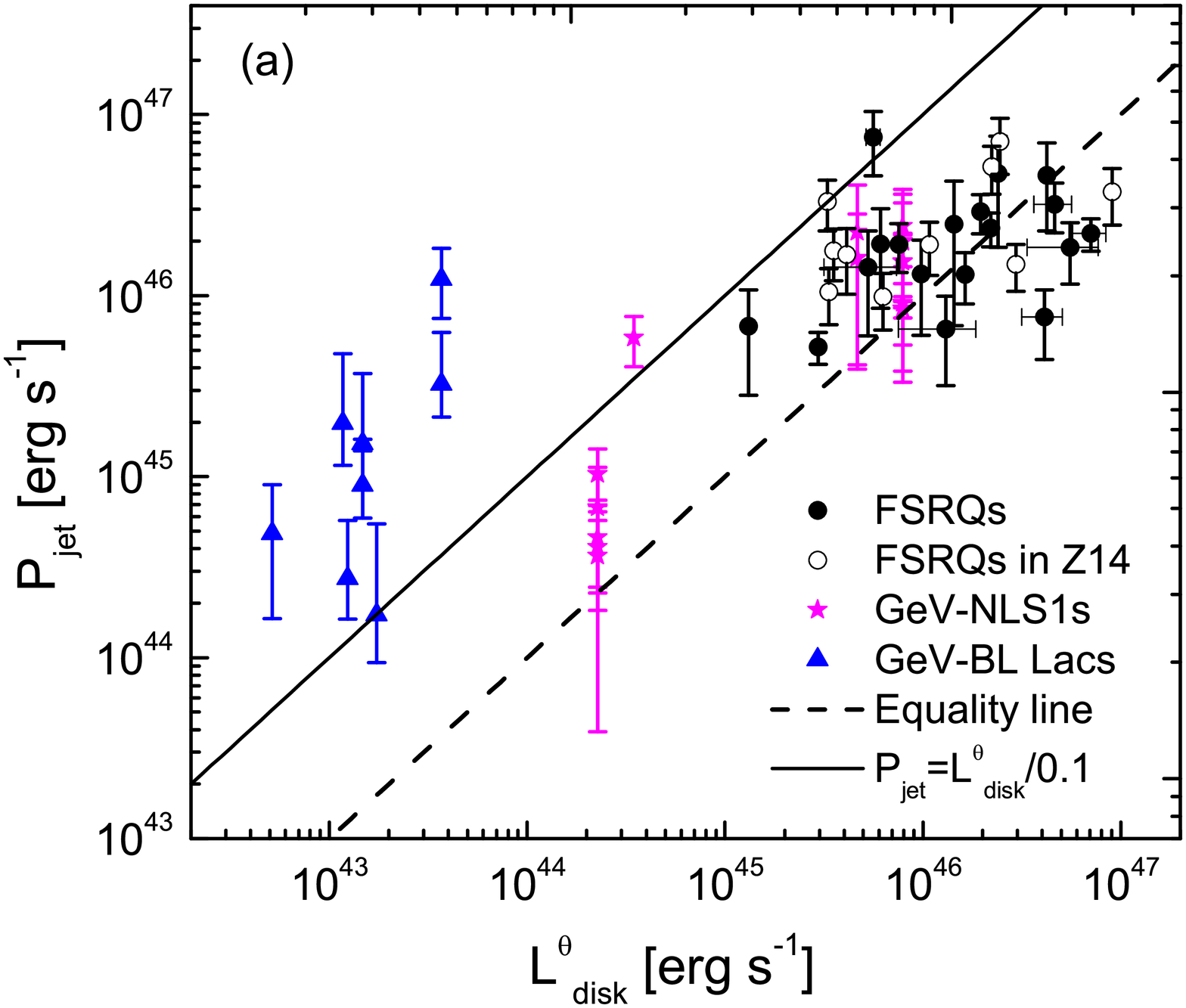}
\includegraphics[angle=0,scale=0.3]{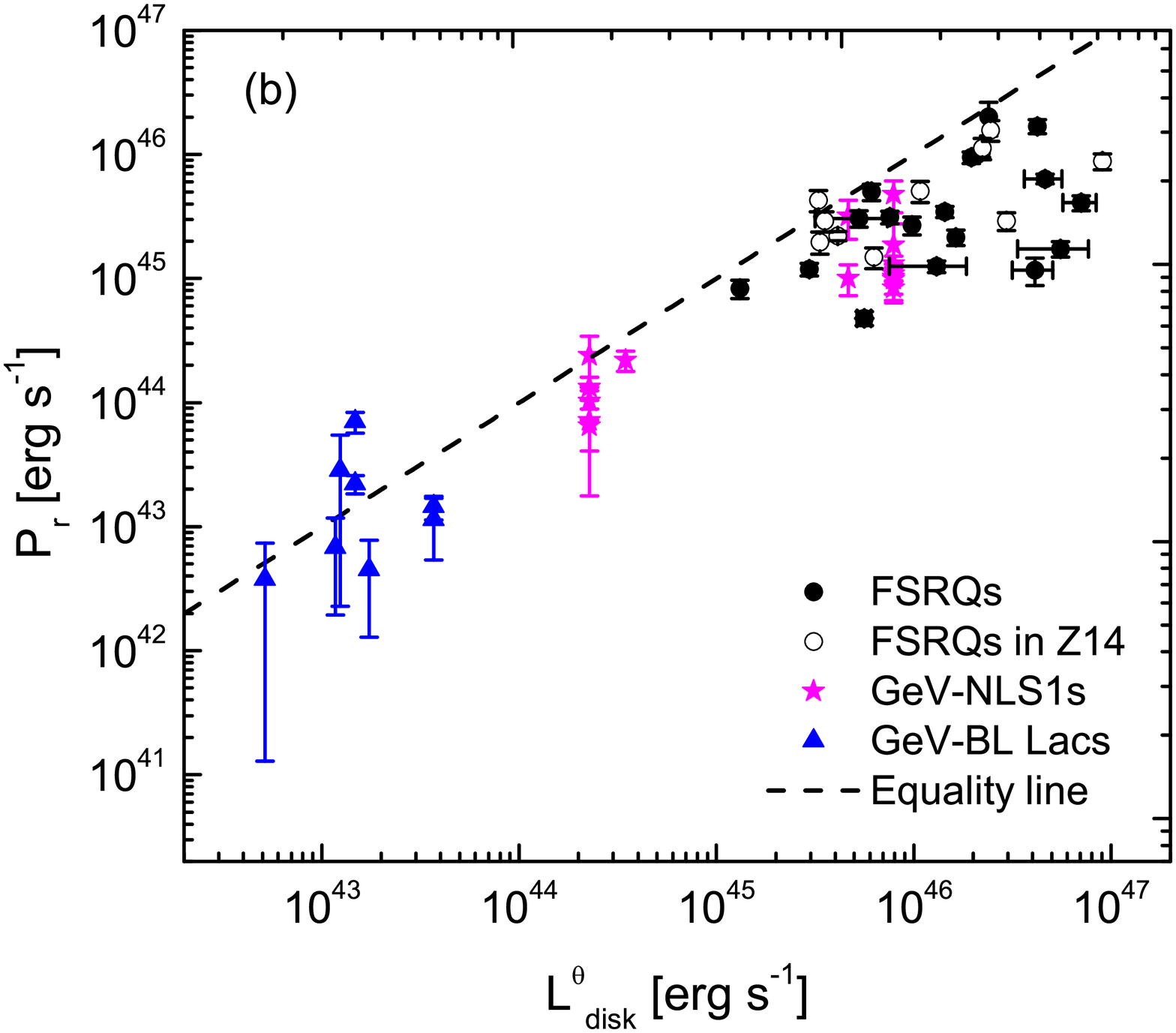}
\caption{$P_{\rm jet}$ and $P_{\rm r}$ as a function of the anisotropic luminosity of accretion disk ($L^{\theta}_{\rm disk}$) for the 18 GeV-FSRQs (black solid circles) in our sample. The data of the 10 GeV FSRQs (black opened circles) in Zhang et al. (2014), the BL Lacs (blue triangles) with available $L_{\rm BLR}$ in Zhang et al. (2012), and the GeV NLS1 galaxies (magenta stars) from Sun et al. (2015) are also presented for comparison.}\label{Pjet-Pr-Ldisk_theta}
\end{figure*}

Recently, Richards \& Lister (2015) reported that the jets of radio-loud (RL) NLS1 galaxies are aligned at moderately small angles to the line of sight, which is similar to blazars. Hence the intrinsic luminosity of accretion disk for GeV NLS1 galaxies should have the same correction factor as FSRQs and BL Lacs, i.e., $L^{\theta}_{\rm disk}=L_{\rm disk}/2$. Taking the anisotropic radiation of accretion disk into account in the $L^{\theta}_{\rm disk}$--$P_{\rm jet}$ and $L^{\theta}_{\rm disk}$--$P_{\rm r}$ planes, the results are given in Figure \ref{Pjet-Pr-Ldisk_theta}. For most of the sources, $P_{\rm jet}$ is higher than $L^{\theta}_{\rm disk}$, but most of the FSRQs and NLS1 galaxies still have higher accretion power of BH ($\dot{M}c^2=L^{\theta}_{\rm disk}/0.1$) than $P_{\rm jet}$, indicating that the pure accretion power is still enough to drive their jet. In the $L^{\theta}_{\rm disk}$--$P_{\rm r}$ plane, the sources still form a sequence and are also closer to the equality line, implying that the radiation power of jet is already comparable with the luminosity of accretion disk for the GeV sources.

Following our previous works, the SED fits in this paper are still under the assumption of $\delta=\Gamma$, i.e., the viewing angle is equal to the opening angle ($1/\Gamma$) of a jet, since the probability is the highest when looking at the jet at the angle of $1/\Gamma$ as described in Section 3.2. It is well known that there is Doppler boosting effect in the radiation of balzars. If the viewing angle is larger than the opening angle of the jet, the leptonic models would not be able to explain the observation data. In another scenario, $\delta=A*\Gamma$ when the viewing angle is smaller than the opening angle of the jet, where the coefficient of $A$ ranges from 1 to 2, the calculated jet powers with the derived parameters would be lower than the previous results (under the assumption of $\delta=\Gamma$). That means the jet powers may be overestimated under the assumption of $\delta=\Gamma$ to fit the SEDs; under this case, the jet powers are further lower than those that derived with the assumption of $\delta=\Gamma$, strengthening the above conclusion that the jet powers are lower than the accretion power of BH.

\subsection{Eddington Ratio and Jet Formation of GeV AGNs}

The radiation of FSRQs and BL Lacs as well as the confirmed GeV NLS1 galaxies (Abdo et al. 2009; Paliya et al. 2013; 2014; Sun et al. 2015; Angelakis et al. 2015) is believed to be dominated by the emission of relativistic jets. The observed spectral properties of the GeV NLS1 galaxies are analogous to GeV FSRQs with significant broad emission lines and the blue bump from accretion disk radiation. BL Lacs are defined as the equivalent width (EW) of emission lines in the rest frame being less than 5 ${\rm \AA}$, and no significant blue bump at $\sim10^{15}$ Hz is observed in their SEDs. In recently years, some sources that are classified as BL Lacs show some properties similar to FSRQs (Sbarufatti et al. 2005; Raiteri et al. 2007; Ghisellini et al. 2011; Giommi et al. 2012), implying that their thermal emission from the accretion disk may be overwhelmed by the jet emisison. Hence the emission line luminosity of the BL Lacs can be used as a proxy to investigate the accretion disk luminosity and then the Eddington ratio of these sources. We collect and compile a sample of $L_{\rm BLR}$ for BL Lacs from literature, which includes 16 BL Lacs with available BH masses as listed in Table 4. Their $L_{\rm disk}$ are also estimated by $L_{\rm BLR}$ with the correlation between $L_{\rm disk}$ and $L_{\rm BLR}$ of the 13 GeV-FSRQs in Figure \ref{corr-Ldisk-Lblr}.

For NLS1 galaxies, besides GeV NLS1 galaxies PMN J0948+0022, PKS 1502+036, and 1H 0323+342, in which their $L_{\rm disk}$ are available and derived by fitting the blue bump in SEDs with the standard accretion disk spectrum (Sun et al. 2015), the other very RL NLS1 galaxies in Yuan et al. (2008) are also taken into account. Using the fluxes of emission lines reported in Yuan et al. (2008), $L_{\rm BLR}$ of these NLS1 galaxies\footnote{For 1H 0323+342, the flux of emission line H$\beta$ is taken from Yao et al. (2015).} are derived with the method proposed in Celotti et al. (1997). The values of $L_{\rm disk}/L_{\rm BLR}$ for the GeV NLS1 galaxies PMN J0948+0022, 1H 0323+342, and PKS 1502+036 are $\sim$86, $\sim$14, $\sim$28, respectively. So $40L_{\rm BLR}$ is used to estimate $L_{\rm disk}$ of other RL NLS1 galaxies in Yuan et al. (2008). It was suggested that the BH masses of NLS1 galaxies derived from the data of the H$\beta$ line may be underestimated (Collin et al. 2006; Zhu et al. 2009; Calderone et al. 2013). Thus the BH masses of RL NLS1 galaxies in Calderone et al. (2013) are used to calculate Eddington ratio. There are 17 very RL NLS1 galaxies with available BH masses in Calderone et al. (2013). Adding 1H 0323+342 (Sun et al. 2015), there are 18 RL NLS1 galaxies as listed in Table 4.

\begin{figure*}
\includegraphics[angle=0,scale=0.5]{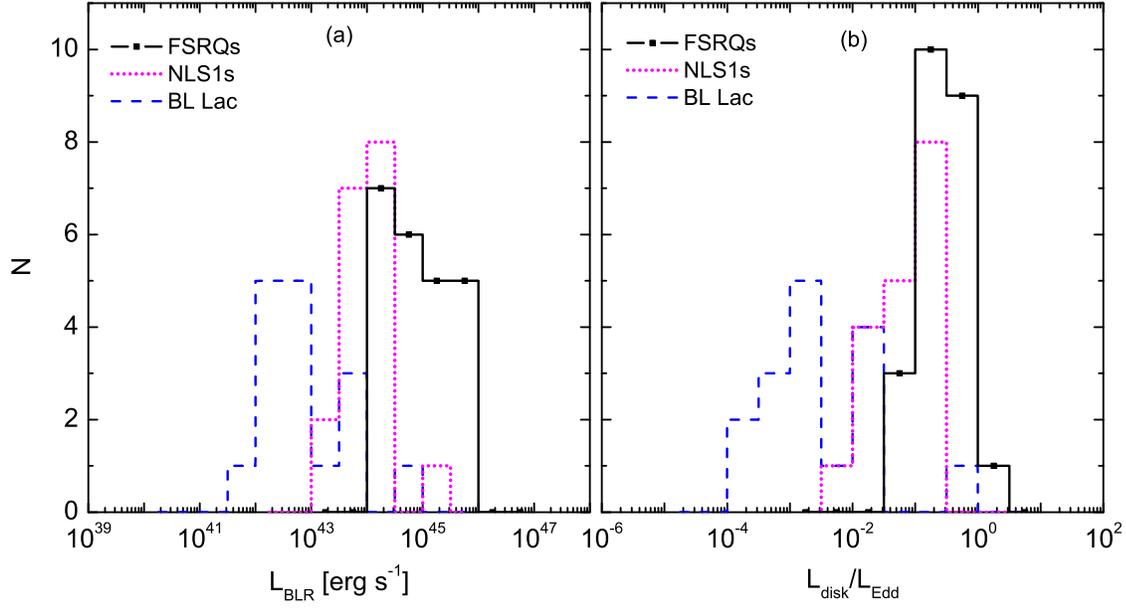}
\caption{Distributions of $L_{\rm BLR}$ and $L_{\rm disk}/L_{\rm Edd}$ for BL Lacs, very RL NLS1 galaxies in Yuan et al. (2008), the GeV FSRQs in our sample and in Zhang et al. (2014).}\label{FSRQ-NLS1-BLLac}
\end{figure*}

The distributions of $L_{\rm BLR}$ and $L_{\rm disk}/L_{\rm Edd}$ for 16 BL Lacs, 18 RL NLS1 galaxies, and 23 GeV FSRQs (including the 10 FSRQs in Z14) are shown in Figure \ref{FSRQ-NLS1-BLLac}. It is found that the distributions of $L_{\rm BLR}$ form a sequence along BL Lac--NLS1--FSRQ, and $L_{\rm BLR}$ of BL Lacs are intrinsically weak compared with that of RL NLS1 galaxies and GeV FSRQs. It may be due to the intrinsically weak radiation of the accretion disk in BL Lacs, and thus it is easily overwhelmed by the jet emission. There is also a BL Lac--NLS1--FSRQ sequence in the $L_{\rm disk}/L_{\rm Edd}$ distribution. It can be found that $L_{\rm disk}/L_{\rm Edd}$ for most of the BL Lacs are lower than 0.01, but $L_{\rm disk}/L_{\rm Edd}$ for the GeV FSRQs is higher than 0.01 (see also Ghisellini et al. 2010; a division to separate BL Lacs and FSRQs). The $L_{\rm disk}/L_{\rm Edd}$ distribution of NLS1 galaxies is intermediate between FSRQs and BL Lacs, and is more analogous to that of FSRQs.

As shown in Figure \ref{Pjet-Pr-Ldisk}, $P_{\rm jet}$ is strongly correlated with $L_{\rm disk}$ and $L_{\rm disk}/L_{\rm Edd}$ if taking all the data of FSRQs, GeV NLS1 galaxies, and BL Lacs into account, indicating the accretion rate is very important to drive jets. Zhang (2013) and Sun et al. (2015) also proposed that the accretion rate (see also Shen \& Ho 2014) may be the fundamental in the unified framework among different types of GeV AGNs; along with the BL Lac--NLS1--FSRQ sequence, the increase of the Eddington ratio corresponds to the change of the accretion disk structure, and then may correspond to the transformation of the dominant mechanism for jet launching. The jets of GeV FSRQs may be powered via extracting the rotational energy of BH with BZ process, which is also consistent with the uncorrelation between $L_{\rm disk}$ and $P_{\rm jet}$ of the GeV FSRQs as described in Section 5. However, the jets of BL Lacs and NLS1 galaxies may be produced via the BP and/or BZ mechanisms, depending on the structure and accretion rate of their accretion disks. Note that the above discussion does not consider the correction of anisotropic radiation of accretion disk mentioned in Section 7.1, but this effect does not change the results since the jets of GeV NLS1 galaxies have the similar properties to that of blazars as discussed in Section 7.1.

\subsection{Implications of unification for RL AGNs}

The topic of the differences and unification among the different types of AGNs has been extensively covered and discussed. According to the unified models for RL AGNs, BL Lacs are associated with center-brightened (FR I) radio galaxies, whereas FSRQs are usually fond in edge-brightened (FR II) radio galaxies (Urry \& Padovani 1995). They may also correspond to the low-excitation radio galaxies (LERGs) and high-excitation radio galaxies (HERGs), although the low/high-excitation radio galaxies do not have one-to-one correspondence with the FR I$-$FR II categories (Hine \& Longair 1979; Laing et al. 1994; Hardcastle et al. 2009). This is also consistent with numerous BL Lacs with FR II radio structure (e.g. Rector \& Stocke 2001) and some quasars with FR I characteristic (e.g., Blundell \& Rawlings 2001). Therefore, intrinsically LERGs and HERGs may be the parent populations of BL Lacs and FSRQs. Recently, some authors suggested that the intrinsic differences between HERGs and LERGs are due to the different modes of accretion, i.e., owing to the different accretion rates (Mingo et al. 2014; Fernandes et al. 2015); HERGs correspond to the radiatively efficient `cold mode' accretion with higher accretion rates, whereas LERGs are the radiatively inefficient `hot mode' accretion with lower accretion rates. The switch lies approximately at the Eddington rates 0.01$\sim$0.04, but the jet production is not switched off. However, the division between the two populations is not very clear, there is still some overlap with wider range of accretion rates for LERGs (Mingo et al. 2014). These results agree well with ours.

As described above, some of the radio properties between HERGs with LERGs do not show clear differences and there is no dependence between jet power with radiative luminosity for HERGs, presumably implying that the process of power conversion into jet power is not solely controlled by the accretion rate; an additional process must be influencing the jet formation, such as the spin of BH (Mingo et al. 2014; Fernandes et al. 2015). This is also consistent with our results that there is no correlation between jet power with Eddington ratio for single population of FSRQs, and the extraction of rotation energy of BH may be important to power their jets.

The large scale kinetic powers of RGs usually are estimated with the radio luminosity at 151 MHz (e.g., Godfrey \& Shabala 2013; Mingo et al. 2014) and they reflect the long-term average power injection from their jets over the whole RL lifetimes of the sources. Averagely, the large kinetic powers of LERGs are lower than that of HERGs (e.g., Mingo et al. 2014). Interestingly, the large scale kinetic powers of blazars is strongly correlated with their jet powers (Zhang et al. 2014), which reflect essentially their instantaneous activities at the present epoch, and the large scale kinetic powers of BL Lacs are lower than that of FSRQs (Zhang et al. 2014; Sun et al. 2015). The statistical associations of BL Lacs with LERGs and FSRQs with HERGs suggest that the current core activities of LERGs also have on average lower power outputs than HERGs.

There are two possible scenarios for the above main differences between the two classes of sources, i.e., BL Lacs/LERGs and FSRQs/HERGs. One scenario is that the two classes are essentially the same sources in nature, but their accretion disks may stay in different accretion states at different times and thus produce different jet/core power outputs. In other words, the two classes of objects would be converted from each other when the accretion states change. This scenario agrees with the fact that their host galaxies seem to show similar properties (Urry \& Padovani 1995; Seymour et al. 2007; Fernandes et al. 2015). However, a main consequence of this scenario is that their long-term average jet/core power outputs should be essentially the same, which contradicts with the observed higher large-scale jet kinetic powers of FSRQs/HERGs than BL Lacs/LERGs. The alternative is that the two classes are essentially different objects with BL Lacs/LERGs tending to produce lower jet/core powers (i.e. with lower accretion rate and/or black hole spin), resulting in the lower large-scale jet kinetic powers in BL Lacs/LERGs. The apparent contradiction in this scenario with the indistinguishable host galaxies of these two classes may be resolved if they have different circum-nuclear environments resulting in different accretion supply on the average. Then the accretion disk instabilities and/or the different jet orientations may result in large scatter of the observed jet/core outputs at different times, in agreement with observations (e.g., Mingo et al. 2014). Future higher quality observations may test if their circum-nuclear environments are indeed different intrinsically.

\section{Summary}

A SED sample of 18 GeV-FSRQs, in which the significant blue bumps from the accretion disk radiation are observed and the BLR luminosities ($L_{\rm BLR}$) of 13 GeV-FSRQs are available, is collected and compiled from literature. On the basis of the SED fits with the jet radiation and accretion disk radiation models, we calculate the jet power ($P_{\rm jet}$) and the accretion disk luminosity ($L_{\rm disk}$), investigate the correlations of the jet power and disk luminosity as well as the BLR luminosity, and also make a comparison of the BLR covering factor ($L_{\rm BLR}/L_{\rm disk}$)  with the large RLQ and RQQ samples from Shen et al. (2011). Our results are summarized below.

\begin{itemize}
\item $L_{\rm BLR}$ of the 13 GeV-FSRQs is strongly correlated with $L_{\rm disk}$, and thus $L_{\rm BLR}$ can be used to estimate $L_{\rm disk}$ when the radiation of accretion disk is overwhelmed by the jet emission. The mean of $L_{\rm BLR}/L_{\rm disk}$ for the 13 GeV-FSRQs is 0.043, which is smaller than that of the RLQs ($\sim$0.082) and RQQs ($\sim$0.076) in Shen et al. (2011). Considering the inclination effect and anisotropic radiation of accretion disk, the $L_{\rm BLR}/L_{\rm disk}$ distribution of the 13 GeV-FSRQs would be same as that of RLQs and RQQs in Shen et al. (2011).

\item No correlation between $P_{\rm jet}$ with $L_{\rm disk}$ and $L_{\rm BLR}$ is found for the GeV FSRQs. $P_{\rm jet}$ for most of the GeV FSRQs is lower than $L_{\rm disk}$ and all of them are lower than the accretion power of BH ($\dot{M}c^2=L_{\rm disk}/0.1$), indicating that the total accretion power of BH is sufficient to drive the jets in these sources, but the uncorrelation between $L_{\rm disk}$ and $P_{\rm jet}$ of the GeV FSRQs in our sample may suggest that their jets are launched by the BZ process via extracting the rotational energy of BH.

\item Using the $L_{\rm BLR}$--$L_{\rm disk}$ relation of the 13 GeV-FSRQs, we estimate $L_{\rm disk}$ of a BL Lac sample with their $L_{\rm BLR}$; a very RL NLS1 galaxy sample from Yuan et al. (2008) is also taken into account. It is found that along with the BL Lac--NLS1--FSRQ sequence $L_{\rm BLR}$ and the Eddington ratio ($L_{\rm disk}/L_{\rm Edd}$) increase, which may correspond to the change of the accretion disk structure and the transformation of the dominant mechanism for jet launching. This is also consistent with the division of their parent populations, i.e., BL Lacs/LERGs and FSRQs/HERGs.
\end{itemize}

\acknowledgments

We thank the anonymous referee for his/her valuable suggestions. We appreciate helpful discussion with Yuan Liu, Wei Cui, Bi-Fang Liu, Wei-Min Yuan, Er-Lin Qiao. This work is supported by the National Basic Research Program (973 Programme) of China (grant 2014CB845800), the National Natural Science Foundation of China (grants 11373036, 11133002, 11025313), the Strategic Priority Research Program ``The Emergence of Cosmological Structures" of the Chinese Academy of Sciences (grant XDB09000000), the Guangxi Science Foundation (2013GXNSFFA019001), and the Young Researcher Grant of National Astronomical Observatories, Chinese Academic of Science. Shuang-Nan Zhang acknowledges support from the Qianren start-up grant 292012312D1117210.

\section{Appendix}

\begin{figure*}
\includegraphics[angle=0,scale=0.3]{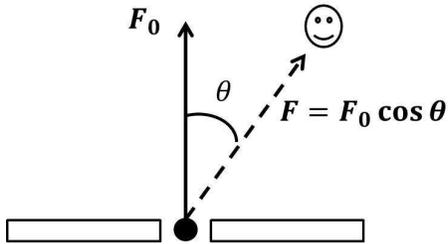}
\caption{Illustration of anisotropic radiation for accretion disk.}\label{Ldisk_theta}
\end{figure*}

As shown in Figure \ref{Ldisk_theta}, the observed flux of accretion disk depends on the inclination angle, i.e., $F=F_0\cos\theta$, where $\theta$ is the angle between the line of sight and the normal of the accretion disk. The intrinsic luminosity of accretion disk is
\begin{equation}
L^{\theta}_{\rm disk}=2D^2_{\rm L}\int_{0}^{2\pi}d\phi\int^{\frac{\pi}{2}}_{0}F_0\cos\theta\sin\theta d\theta=2\pi D^2_{\rm L}F_0,
\end{equation}
where $D_{\rm L}$ is the luminosity distance of the source. The derived isotropic luminosity is
\begin{equation}
L_{\rm disk}=4\pi D^2_{\rm L}F=4\pi D^2_{\rm L}F_0\cos\theta.
\end{equation}
Therefore, the inferred intrinsic luminosity should have a correction factor with the derived isotropic luminosity, i.e.,
\begin{equation}
L^{\theta}_{\rm disk}=\frac{L_{\rm disk}}{2\cos\theta}.
\end{equation}

\begin{deluxetable}{lccccc}
\centering
\tabletypesize{\footnotesize}\tablecolumns{6}\tablewidth{28pc} \tablecaption{The data for BLRs of 18 GeV-FSRQs in our sample}\tablenum{1}
 \tablehead{\colhead{Source}  &  \colhead{Flux$_{\rm H\beta}$\tablenotemark{a}} &  \colhead{Line\tablenotemark{b}}& \colhead{$\log L_{\rm BLR}$\tablenotemark{a}}  &  \colhead{$R_{\rm BLR}$\tablenotemark{c}} & \colhead{$U_{\rm BLR}$\tablenotemark{c}}}
\startdata

3C 273&1548&H$\beta$$^{\rm C97}$&45.50&238.3&3.13\\
3C 454.3&43&H$\beta$$^{\rm C97}$&45.57&287.0&2.51\\
PKS 0208$-$512&12&Mg~{\scriptsize II}$^{\rm C97}$&45.17&190.9&2.27\\
PKS 0420$-$01&8&Mg~{\scriptsize II}$^{\rm C97}$&44.88&142.9&2.07\\
PKS 0528+134&1.5&C~{\scriptsize IV}$^{\rm P11}$&45.06&168.3&2.28\\
B3 0650+453&\nodata&\nodata&44.24&70.6&1.95\\
PKS 0727$-$11&\nodata&\nodata&\nodata&\nodata&\nodata\\
PKS 1127$-$145&19.36&Ly$\alpha$$^{\rm T12}$&45.75&287.4&3.82\\
1Jy  1308+326&3.98&H$\beta$$^{\rm C09}$ &44.99&114.4&4.20\\
PKS 1508$-$055&\nodata&\nodata&45.50&267.3&2.48\\
PKS 1510$-$089&53&H$\beta$$^{\rm C97}$ &44.72&119.9&2.04\\
TXS 1846+322&\nodata&\nodata&44.56&98.7&2.07\\
PKS 2123$-$463&\nodata&\nodata&\nodata&\nodata&\nodata\\
TXS 2141+175&20.95&C~{\scriptsize IV}$^{\rm O94}$&44.23&45.0&4.66\\
PKS 2142$-$758&\nodata&\nodata&\nodata&\nodata&\nodata\\
PKS 2144+092&\nodata&\nodata&\nodata&\nodata&\nodata\\
PKS 2204$-$54&\nodata&\nodata&\nodata&\nodata&\nodata\\
PMN 2345$-$1555&\nodata&\nodata&44.33&78.1&1.98\\

\enddata
\tablenotetext{a}{The flux of emission line H$\beta$ is in units of $10^{-15}$ erg s$^{-1}$ cm$^{-2}$. $L_{\rm BLR}$ is in units of erg s$^{-1}$. The luminosities of line H$\beta$ for B3 0650+453, PKS 1508$-$055, and PMN 2345$-$1555 are derived with $L_{\rm BLR}$ when it is used to estimate $R_{\rm BLR}$.}
\tablenotetext{b}{The line used to estimate the flux of emission line H$\beta$ with the relative flux ratios. The superscripts denote the references--C97: Celotti et al. (1997); P11: Palma et al. (2011); T12: Tang et al. (2012); C09: Chen et al. (2009); O94: Osmer et al. (1994).}
\tablenotetext{c}{$R_{\rm BLR}$ and $U_{\rm BLR}$ are in units of light days and $10^{-2}$ erg cm$^{-3}$, respectively.}
\end{deluxetable}

\begin{deluxetable}{lccccccccccccl}
\tabletypesize{\tiny} \rotate \tablecolumns{14}\tablewidth{48pc} \tablecaption{Derived Parameters from Our SED Fits with the Single-zone Leptonic Model}\tablenum{2}
\tablehead{\colhead{Source}  & \colhead{$z$}  & \colhead{$p_1$} &  \colhead{$p_2$} &  \colhead{$\delta$} & \colhead{$B$}   & \colhead{$\gamma_{\min}$} & \colhead{$\gamma_{\rm b}$} &\colhead{$\gamma_{\rm max}$} & \colhead{$N_{0}$} & \colhead{$\log L_{\rm jet}$\tablenotemark{a}} &  \colhead{$\log L_{\rm disk}$} &  \colhead{$\log P_{\rm jet}$} & \colhead{$M_{\rm BH}$\tablenotemark{b}} \\
\colhead{}  & \colhead{}  & \colhead{} & \colhead{} &  \colhead{} &\colhead{[G]} & \colhead{} & \colhead{} & \colhead{}
& \colhead{[$\times10^{3}$ cm$^{-3}$]}&  \colhead{[erg s$^{-1}$]}& \colhead{[erg s$^{-1}$]}& \colhead{[erg s$^{-1}$]}&\colhead{[$\log M_{\bigodot}$]}}
\startdata

3C 273&0.158&1.4&4.0&7.4$\pm$0.9&8.5$\pm$1.6&6$^{+19}_{-5}$&328$\pm$79&2000&5.6$\pm$5.3&47.11$\pm$0.03&46.92$\pm$0.10&45.88$\pm$0.18&9.3$^{\rm F04}$\\
3C 454.3&0.859&1.2&3.46&15.6$\pm$0.6&5.1$\pm$0.8&40$\pm$6&137$\pm$32&1800&3.4$\pm$2.0&48.49$\pm$0.02&46.97$\pm$0.09&46.50$\pm$0.13&9.17$^{\rm W02}$\\
PKS 0208$-$512&1.003&1.3&3.56&15.8$\pm$0.7&3.4$\pm$1.2&40$\pm$13&105$\pm$40&3300&6.9$\pm$6.9&48.24$\pm$0.02&46.46&46.39$\pm$0.32&9.21$^{\rm F04}$\\
PKS 0420$-$01&0.916&2.64&4.06&12.8$\pm$0.7&8.2$\pm$1.1&51$\pm$11&385$\pm$118&5000&2512$\pm$1388&48.01$\pm$0.02&46.18&46.28$\pm$0.13&9.03$^{\rm W02}$\\
PKS 0528+134&2.07&1.2&4.2&17.4$\pm$0.9&2.9$\pm$1.1&111$\pm$6&230$\pm$86&7000&16.6$\pm$16.6&49.01$\pm$0.03&46.93&46.66$\pm$0.22&9.4$^{\rm X04}$\\
B3 0650+453&0.933&1.28&3.6&14.1$\pm$1.0&1.3$\pm$0.3&95$\pm$16&111$\pm$37&4000&15.8$\pm$15.8&48.08$\pm$0.02&46.02$\pm$0.17&46.16$\pm$0.25&8.17$^{\rm S12}$\\
PKS 0727$-$11&1.589&1.8&3.9&20.6$\pm$1.2&5.4$\pm$1.1&50$\pm$12&254$\pm$61&5000&15.8$\pm$10.6&48.91$\pm$0.03&46.59&46.46$\pm$0.11&\nodata\\
PKS 1127$-$145&1.184&1.2&4.56&13.1$\pm$0.8&11.5$\pm$1.2&49$\pm$4&213$\pm$32&4000&3.7$\pm$2.1&48.15$\pm$0.03&47.15$\pm$0.08&46.35$\pm$0.09&9.18$^{\rm Z09}$\\
1Jy  1308+326&0.997&2.2&3.4&12.6$\pm$0.9&3.4$\pm$0.9&44$\pm$10&353$\pm$129&8000&490$\pm$451&48.20$\pm$0.02&46.09&46.29$\pm$0.24&8.94$^{\rm C09}$\\
PKS 1508$-$055&1.185&1.2&3.8&17.0$\pm$1.1&7.0$\pm$1.5&32$\pm$14&141$\pm$41&5000&0.9$\pm$0.7&48.00$\pm$0.03&47.04$\pm$0.17&46.27$\pm$0.16&8.97$^{\rm Z09}$\\
PKS 1510$-$089&0.36&1.12&3.72&11.0$\pm$0.5&3.1$\pm$0.5&20$\pm$12&305$\pm$32&1500&0.4$\pm$0.2&47.46$\pm$0.03&45.77&45.72$\pm$0.09&8.65$^{\rm W02}$\\
TXS 1846+322&0.798&1.4&3.76&13.1$\pm$0.6&2.5$\pm$0.7&69$\pm$8&206$\pm$72&5000&4.9$\pm$4.4&47.63$\pm$0.03&46.42$\pm$0.18&45.82$\pm$0.22&8.21$^{\rm S12}$\\
PKS 2123$-$463&1.67&1.3&3.8&17.9$\pm$0.6&3.6$\pm$0.6&88$\pm$5&243$\pm$56&5000&4.8$\pm$2.8&48.82$\pm$0.02&46.65&46.37$\pm$0.09&\nodata\\
TXS 2141+175&0.213&1.4&4.02&10.3$\pm$0.6&5.1$\pm$0.8&2$\pm$1&42$\pm$6&8000&15$\pm$8.1&47.01$\pm$0.02&46.05$\pm$0.03&46.87$\pm$0.17&8.98$^{\rm F03}$\\
PKS 2142$-$758&1.139&1.6&3.6&9.5$\pm$1.4&1.7$\pm$1.0&197$\pm$17&422$\pm$188&3000&856$\pm$856&48.56$\pm$0.03&46.68&46.67$\pm$0.26&\nodata\\
PKS 2144+092&1.113&1.4&4.0&14.3$\pm$1.0&3.8$\pm$1.2&74$\pm$8&175$\pm$66&4000&11.5$\pm$11.5&48.04$\pm$0.03&46.29&46.12$\pm$0.24&\nodata\\
PKS 2204$-$54&1.215&1.4&4.4&14.4$\pm$0.9&5.7$\pm$1.2&52$\pm$5&205$\pm$46&4000&6.5$\pm$4.4&47.95$\pm$0.03&46.52&46.12$\pm$0.14&\nodata\\
PMN 2345$-$1555&0.621&2.0&4.0&13.8$\pm$1.0&2.6$\pm$0.8&46$\pm$11&141$\pm$55&8000&9.3$\pm$9.3&47.50$\pm$0.04&45.42&45.83$\pm$0.25&8.16$^{\rm S12}$\\

\enddata
\tablenotetext{\rm a}{The errors of jet radiation luminosity are estimated with the errors of the two peak luminosities only.}
\tablenotetext{\rm b}{BH masses of the sources. The superscripts denote the references--F04: Fan \& Cao (2004); W02: Woo \& Urry (2002); X04: Xie et al. 2004; S12: Shaw et al. (2012); Z09: Zhou \& Cao; C09: Chen et al. (2009); and F03: Falomo et al. (2003).}

\end{deluxetable}

\begin{deluxetable}{lccc}

\tabletypesize{\footnotesize}\tablecolumns{4}\tablewidth{40pc} \tablecaption{K$-$S Test Probability of $p_{\rm KS}$\tablenotemark{a} for the Parameter Distributions among the 13 GeV-FSRQs, RLQs, and RQQs}\tablenum{3}
\tablehead{\colhead{Parameters\tablenotemark{b}} & \colhead{FSRQs \& RLQs} & \colhead{FSRQs \& RQQs} & \colhead{RLQs \& RQQs}}
\startdata
$L_{\rm disk}$--$L_{\rm BLR}$& 0.045&0.29&1.3E-24\\
$L_{\rm BLR}$&0.38&0.59&4.9E-12\\
$L_{\rm disk}$&0.06&0.07&7.1E-7\\
$L_{\rm BLR}/L_{\rm disk}$&2.3E-4&0.002&3.9E-11\\
$M_{\rm BH}$&0.13&0.45&1.5E-27\\
$L_{\rm disk}/L_{\rm Edd}$&2.0E-5&0.015&6E-37\\
\enddata
\tablenotetext{\rm a}{$p_{\rm KS}$ larger than 0.1 (0.2 for two-dimensional K$-$S test) would strongly suggest no statistical difference between two samples, whereas $p_{\rm KS}$ smaller than 10$^{-4}$ would suggest there is significantly statistical difference between two samples.}
\tablenotetext{\rm b}{The two-dimensional K$-$S test is only used for the $L_{\rm disk}$--$L_{\rm BLR}$ relation, and the one-dimensional K$-$S tests are given for the other parameter distributions between two samples.}
\end{deluxetable}

\begin{deluxetable}{lccccccccccc}
\tabletypesize{\tiny} \rotate \tablecolumns{14}\tablewidth{43pc} \tablecaption{BLR Luminosities and BH Masses of RL AGNs}\tablenum{4}
\tablehead{\multicolumn{4}{c}{RL NLS1 galaxies\tablenotemark{a}}&\multicolumn{4}{c}{BL Lacs\tablenotemark{b}}&\multicolumn{4}{c}{GeV FSRQs in Z14\tablenotemark{c}}\\
\cline{1-4}\cline{5-8}\cline{9-12}\\
\colhead{Source}  & \colhead{$z$} &\colhead{$\log L_{\rm BLR}$} & \colhead{$M_{\rm BH}$} & \colhead{Source}  & \colhead{$z$} &\colhead{$\log L_{\rm BLR}$} & \colhead{$M_{\rm BH}$} & \colhead{Source}  & \colhead{$z$} &\colhead{$\log L_{\rm BLR}$} & \colhead{$M_{\rm BH}$}\\
\colhead{} & \colhead{} & \colhead{[erg s$^{-1}$]}&\colhead{[$\log M_{\bigodot}$]} & \colhead{} & \colhead{} & \colhead{[erg s$^{-1}$]}&\colhead{[$\log M_{\bigodot}$]}& \colhead{} & \colhead{} & \colhead{[erg s$^{-1}$]}&\colhead{[$\log M_{\bigodot}$]}}

\startdata
1H0323+342&0.0629&43.53&8.6&PKS 0521$-$36&0.055&42.68&8.6&S4 0133+47&0.859&44.42&8.3\\
J0814+5609&0.509&44.28&8.4&PKS 0829+46&0.174&42.57&8.68&4C 28.07&1.213&45.30&9.22\\
J0850+4626&0.523&43.93&8.8&PKS 0851+202&0.306&42.83&8.86&PKS 0454$-$234&1.003&44.39&9.17\\
J0902+0443&0.532&44.09&8.5&TXS 0954+658&0.367&42.45&8.53&S4 0917+44&2.19&45.77&9.29\\
J0948+0022&0.584&44.03&9.1&PMN 1012+063&0.727&42.89&8.5&4C 29.45&0.729&44.66&8.61\\
J0953+2836&0.657&44.01&8.3&PKS 1057$-$79&0.581&43.76&8.8&3C 279&0.536&44.48&8.28\\
J1037+0036&0.595&43.78&9.1&Mkn 421&0.031&41.69&8.67&PKS 1502+106&1.839&45.23&8.98\\
J1146+3236&0.465&44.11&8.8&1ES 1218+304&0.182&42.06&8.58&B2 1520+31&1.487&44.88&8.92\\
J1238+3942&0.622&43.76&8.2&W Com&0.102&42.14&\nodata&4C 66.20 &0.657&44.40&9.14\\
J1246+0238&0.362&43.77&8.7&PKS 1519$-$273&1.294&43.53&8.8&PKS 2325+093&1.843&45.19&8.7\\
J1305+5116&0.785&45.17&9.2&Mkn 501&0.034&42.20&9.03&&&&\\
J1435+3131&0.501&43.98&8.8&PKS 1749+096&0.322&43.70&8.66&&&&\\
J1443+4725&0.703&44.30&8.4&S5 1803+78&0.68&44.85&8.36&&&&\\
J1505+0326&0.408&43.39&8.3&TXS 1807+698&0.051&42.00&8.7&&&&\\
J1548+3511&0.478&44.26&8.7&PKS 2005$-$489&0.071&42.04&8.76&&&&\\
J1634+4809&0.494&43.86&8.9&BL Lacertae&0.069&42.52&8.61&&&&\\
J1644+2619&0.144&43.14&8.2&PKS 2240$-$260&0.774&43.46&8.6&&&&\\
J1722+5654&0.425&44.06&8.1&&&&&&&&\\
\enddata
\tablenotetext{\rm a}{$L_{\rm BLR}$ is estimated with the fluxes of emission lines in Yuan et al. (2008). The BH masses of these sources are taken from Calderone et al. (2013).}
\tablenotetext{\rm b}{The data are taken from Ghisellini et al. (2011) and Sbarrato et al. (2012).}
\tablenotetext{\rm c}{The values of $L_{\rm BLR}$ and $M_{\rm BH}$ are taken from Xiong \& Zhang (2014). $L_{\rm BLR}$ have been corrected for the different cosmological parameters.}
\end{deluxetable}

\clearpage

\end{document}